\documentclass[a4paper,fleqn,usenatbib,useAMS]{mnras}
\usepackage{amsmath,fleqn,graphicx,txfonts}
\usepackage{times}
\usepackage{gensymb}
\usepackage{epsf}
\usepackage[dvips]{epsfig}
\usepackage{xspace}
\usepackage{amssymb}
\usepackage[british,UKenglish]{babel}
\usepackage{multicol, blindtext} 
\usepackage{bm}
\usepackage{pdflscape}

\newcommand{\um}{\ensuremath{\,{\rm u_M}}\xspace}
\newcommand{\ud}{\ensuremath{\,{\rm u_d}}\xspace}
\newcommand{\uv}{\ensuremath{\,{\rm u_v}}\xspace}
\newcommand{\ut}{\ensuremath{\,{\rm u_t}}\xspace}
\newcommand{\e}{\ensuremath{\,{\!=\!}\,}\xspace}
\newcommand{\sm}{\ensuremath{\,{\rm M_{\odot}}}\xspace}
\newcommand{\slu}{\ensuremath{\,{\rm L_{\odot}}}\xspace}
\newcommand{\mum}{\ensuremath{\,{\rm \mu m}}\xspace}
\newcommand{\si}{\ensuremath{\,{\!\sim\!}\,}\xspace}
\newcommand{\ie}{{\it i.e.}\xspace}

\newcommand{\kpc}{\ensuremath{\,{\rm kpc}}\xspace}
\newcommand{\pc}{\ensuremath{\,{\rm pc}}\xspace}
\newcommand{\Myr}{\ensuremath{\,{\rm Myr}}\xspace}
\newcommand{\Gyr}{\ensuremath{\,{\rm Gyr}}\xspace}
\newcommand{\kms}{\ensuremath{\,{\rm km}\,{\rm s}^{-1}}\xspace}

\newcommand{\K}{\ensuremath{{K_s}}\xspace}

\newcommand{\as}{\ensuremath{\,{\rm arcsec}}\xspace}

\begin{document}
\title[Dynamical Models for M31 - Bulge \& Bar]
{Andromeda chained to the Box -- Dynamical Models for M31: Bulge \& Bar}

\author[Bla\~{n}a et al.]
{Matias Bla\~{n}a D\'iaz$^{1}$\thanks{E-mail: mblana@mpe.mpg.de},
Christopher Wegg$^{1}$, 
Ortwin Gerhard$^{1}$, 
Peter Erwin$^{1,2}$, 
Matthieu Portail$^{1}$,
\newauthor 
Michael Opitsch$^{1,2,3}$,
Roberto Saglia$^{1,2}$,
Ralf Bender$^{1,2}$\\
$^{1}$ Max-Planck-Institut f\"ur extraterrestrische Physik, Gie\ss enbachstra\ss e 1, 85748 Garching bei M\"unchen, Germany\\
$^{2}$ Universitäts-Sternwarte M\"unchen, Scheinerstr. 1, D-81679 M\"unchen, Germany\\
$^{3}$ Exzellenzcluster Universe, Boltzmannstr. 2, D-85748 Garching bei M\"unchen, Germany}

\pagerange{\pageref{firstpage}--\pageref{lastpage}} \pubyear{2016}
\maketitle
\label{firstpage}

\begin{abstract}
  Andromeda is our nearest neighbouring disk galaxy and a prime
  target for detailed modelling of the evolutionary processes that
  shape galaxies.  We analyse the nature of
  M31's triaxial bulge with an extensive set of N-body models, which
  include Box/Peanut (B/P) bulges as well as initial classical bulges
  (ICBs). Comparing with IRAC 3.6\mum data, only one model
  matches simultaneously all the morphological properties of M31's
  bulge, and requires an ICB and a B/P bulge with 1/3 and 2/3 of
  the total bulge mass respectively.  We find that our pure B/P bulge models do not
  show concentrations high enough to match the S\'ersic index ($n$) and 
  the effective radius of M31's bulge. Instead, the best model requires an ICB component
  with mass $M^{\rm ICB}\e1.1\times10^{10}\sm$ and
  three-dimensional half-mass radius $r_{\rm half}^{\rm ICB
  }\e0.53\kpc\,(140\as)$.  The B/P bulge component has a mass of $M^{\rm
    B/P}\e2.2\times10^{10}\sm$ and a half-mass radius of
  $r_{\rm half}^{\rm B/P}\e1.3\kpc\,(340\as)$. 
  The model's B/P bulge extends to $r^{\rm B/P}\e3.2\kpc\,(840\as)$ 
  in the plane of the disk, as does M31's bulge. 
  In this composite bulge
  model, the ICB component explains the velocity dispersion drop
  observed in the centre within $R\!<190\pc\,(50\as)$, while the B/P
  bulge component reproduces the observed rapid rotation and 
  the kinematic twist of the observed zero velocity line.   
  This model's pattern speed is
  $\Omega_p\e38\kms\kpc^{-1}$, placing corotation at
  $r_{\rm cor}\e5.8\kpc\,(1500\as)$. The outer Lindblad resonance
  (OLR) is then at $r_{\rm OLR}\e10.4\kpc$, near the 10\kpc-ring of
  M31, suggesting that this structure may be related to the
  bar's OLR. 
  By comparison with an earlier snapshot, we estimate that M31's thin bar extends to 
  $r_{\rm bar}^{\rm thin}\si4.0\kpc\,(1000\as)$ in the disk plane,
 and in projection extends to $R_{\rm bar}^{\rm thin}\si2.3\kpc\,(600\as)$.
 \end{abstract}


\section{Introduction}
\label{sec:intro}
The Andromeda galaxy (M31, NGC224) is the nearest spiral galaxy, located at a distance of $785\pm25\,{\rm kpc}$
from the Milky Way (MW) \citep{McConnachie2005}, which makes it one of the best cases to study in 
great detail the structural and evolutionary properties of spiral galaxies.
It shows several signs of a hierarchical formation history through its satellites, the Giant 
Stream and its old stellar halo suggesting a history of minor mergers \citep{Ibata2001, Tanaka2010} 
or also a possible major merger \citep{Hammer2010, Bekki2010}. It also shows signs of secular 
evolution such as a massive star forming disk, spiral arms and rings \citep{Gordon2006, Barmby2006, Chemin2009}. 
The massive bulge of Andromeda has commonly been considered in the literature as a classical 
bulge, or more recently, as a classical bulge with ``pseudobulge trimmings'' \citep{Mould2013}.

According to theory, classical bulges are considered as elliptical galaxies in the 
centres of disks, as remnants of a very early formation process, or as remnants of mergers of 
galaxies occurred during the first gigayears  
of a violent hierarchical formation \citep{Toomre1977, Naab2003, 
Bournaud2005, Athanassoula2016}. Another proposed mechanism to form early bulges consists of
instabilities in the early disk, forming clumps and star clusters that spiral in due to 
dynamical friction, merging later and forming a bulge \citep{Noguchi1999, Immeli2004, Bournaud2007}. 
The comprehensive review of \citet{Kormendy2013} describes in general how disc galaxies can also 
develop bulges through secular evolution, referred also as pseudobulges. 
The term pseudobulge groups disky bulges and box/peanut bulges together. 
In this scenario pseudobulges are a manifestation of the evolution of the disk, contrary to the classical bulges.

\citet{Fisher2008} use a sample 
of spiral galaxies and their photometric properties such as the S\'ersic index $n$, the effective 
radius $R_{\rm e}$, and the disk scale length $R_d$, to distinguish pseudobulges from classical bulges. 
They conclude that usually pseudobulges have 
S\'ersic indices $n\!\lesssim\!2$ and classical bulges indices $n\!\gtrsim\!2$.
\citet{Fabricius2012} additionally show a correlation between the 
S\'ersic index, the velocity dispersion and the rotation. They compare the S\'ersic index and 
the velocity dispersion averaged within one tenth of the effective radius $\sigma_{R_{\rm e}/10}$. 
Their sample shows that classical bulges, defined morphologically or through 
a S\'ersic index of $n\!\gtrsim\!2.1$, tend to have higher $\sigma_{R_{\rm e}/10}$, rarely getting 
lower than $100\,$km s$^{-1}$ and with a mean of the sample of $\si\!150\,$km s$^{-1}$. 
Pseudobulges, defined with $n\!\lesssim\!2.1$, show lower $\sigma_{R_{\rm e}/10}$, some as low as 
$\si\!50\,$km s$^{-1}$, and a mean value of $100\,$km s$^{-1}$ for the sample. 
However these classification criteria have only statistical meanings, as a particular galaxy, and its bulge,
may actually present properties of both types of bulges.

Simulations of wet mergers by \citet{Keselman2012} showed that classical or primordial 
bulges may also present characteristics of pseudobulges, such as rotation and $n<2$, which makes even 
harder to distinguish the origin of its properties. 
Furthermore, \citet{Erwin2014} showed with observations that classical bulges 
can coexist with disky pseudobulges and box/peanut bulges, a situation 
that has also been reproduced in simulations \citep{Athanassoula2016}.

\citet{Saha2012} showed that  if a low mass non-rotating classical 
bulge is present during the formation and evolution of the bar it absorbs angular 
momentum from the bar \citep[see also][]{Hernquist1992, Athanassoula2003}, and subsequently the bulge
manifests cylindrical rotation like the bar, thereby making the final 
combined structure with the bar difficult to disentangle. However, more massive classical bulges do not reach 
cylindrical rotation \citep{Saha2016}.

Observations show that M31's bulge is triaxial, with a photometric twist relative to the 
disk \citep{Lindblad1956, Hodge1982, Beaton2007}, and the misalignment of the bulge kinematic major 
axis and its photometric major axis \citep{Saglia2010}.
A known dynamical mechanism to generate a triaxial structure is the buckling instability 
of the bar, which generates the box/peanut (B/P) bulge in N-body models 
\citep[see Section \ref{sec:meth:sim:bar}]{Combes1990, Raha1991} . 
 \citet{Erwin2016} have recently found indications that two 
barred galaxies (NGC 3227 and NGC 4569) are currently in the buckling phase.

Several hints such as: the kinematics of the bulge, the 
boxiness of the surface-brightness contours, and the twist of the photometric 
major axis of the bulge's boxy region and the disk's major axis, 
suggest that M31 could contain a B/P bulge and 
therefore may also host a thin bar. The inclination of $\si77^{\rm o}$ \citep{Corbelli2010} 
of the disk presents challenges, but the proximity of M31 and the excellent available 
photometric and kinematic data \citep{Barmby2006, Chemin2009, Corbelli2010, Saglia2010}, 
make it possible to explore this scenario.
\citet{Courteau2011} (hereafter Co11) produced photometric models from the IRAC 3.6\mum data and find 
a bulge-to-total luminosity ratio of $(B/T)_{3.6\mum}\si0.3$ a S\'ersic index of 
$n\e2.4\pm0.2$, an effective radius $R_{\rm e}\e1.10\pm0.10\kpc$ or 
$\approx\!\!290\,\as$  (M31 distance implies a unit conversion of 
$3.8\pc\e1\as$,  $1\kpc\e260\as$ and $13.7\kpc\e1\degree$), and a 
disk scale length of $R_d\e5.8\pm0.1\kpc$, giving a ratio of $R_{\rm e}/R_d\si0.19$.
These values would locate M31's bulge just between the classical and the pseudobulges, 
according to \citet{Fisher2008}. However, the kinematics in the bulge region 
\citep[see the major axis velocity and dispersion profiles in their Fig.3]{Saglia2010}, would 
classify it among the group of classical bulges \citep{Fabricius2012}, due to the high velocity 
dispersion at $R_{\rm e}/10$ of $\sigma_{R_{\rm e}/10}\!\!\gtrsim\!\!150\,$km s$^{-1}$ and the rather slow rotation 
in the bulge region ($\si70\,$km s$^{-1}$ at $R_{\rm e}$) .

Another interesting hint of the bulge type of M31 comes from the metallicity analysis and 
the age estimation of the stars. \citet{Saglia2010} find the presence of 
a very old stellar population in the central part within $300\as$,  
with an average age of \si12\Gyr, which would support the idea of a classical component in the 
bulge. 
Meanwhile, from $360\as$ outwards the mean age of the stellar population drops to $<$8 Gyr. 
This is compatible with the isophotal analysis performed by \citet{Beaton2007}, where the authors 
present strong evidence for the presence and projected extension of the B/P bulge in M31, by quantifying 
the morphological properties of the isophotes of M31 2MASS 6X images in the J,H,K$_{\rm s}$ 
bands.
Within $\si50\as$ the round isophotes resemble a classical bulge with low ellipticity. Beyond  $\si50\as$ 
the boxy isophotes emerge, also increasing the ellipticity, extending until $\si700\as$, where the strongest 
boxiness is at $\si300\as$. \citet[hereafter AB06]{Athanassoula2006} used four N-body models 
of barred galaxies based on \citet{Athanassoula2003}, two with a pure B/P bulge and two with
a classical bulge combined with a B/P bulge, to compare qualitatively the shape of their isophotes 
with the isophotes of M31 in the J band, concluding the presence of a dominant B/P bulge and 
also a classical bulge component in M31. 

With the IRAC 3.6\mum 
data that are less affected by dust absorption than the 2MASS 6X data and go deeper into the 
bulge and disk, together with the new kinematic data of \citet{Saglia2010, Chemin2009, Corbelli2010} and 
Opitsch et al. (in preparation), we can quantitatively compare M31 with  $N$-body models. 
From these observational quantities we can build an understanding of the inner structure of M31's bulge 
and estimate the contribution of each possible subcomponent \ie 
a classical bulge, a B/P bulge and a thin bar.
The goal of this paper is to explore dynamical models for the bulge of M31 that consist 
of two components, a  classical component and a B/P bulge, and also test pure B/P bulge models.

The paper is ordered as follows: 
Section \ref{sec:obsm31} characterises the observational data, 
Section \ref{sec:meth} describes the set-up of the simulations, 
and our technique to compare the simulations with the observations.
Section \ref{sec:res} shows simultaneously the results for M31 and 
for the simulations. The morphological and photometric analysis are presented in Sections 
\ref{sec:res:morph} and \ref{sec:res:phot}. 
In Section \ref{sec:res:bestmod} we present the properties of the best model. 
In Section \ref{sec:bulgeM31} we discuss properties of triaxial models in the literature,
and in Section \ref{sec:conc} we conclude with a summary of our findings and the
implications for M31.

\section{Observational data: M31 IRAC 3.6\mum image}
\label{sec:obsm31}
The imaging data we use for our analysis come from the large-scale
IRAC mosaic images of the \textit{Spitzer Space Telescope} \citep{Barmby2006} -- specifically,
the 3.6\mum IRAC1 mosaic shown in Fig.\ref{fig:SBmapM31}, kindly made available to us by Pauline
Barmby. The near side of the disk is in the upper part of the figure, pointing north-west, as 
evidenced by the dust lanes and reddening maps \citep{Walterbos1988}.
Careful inspection of the edges of the mosaic showed extended regions
with slightly negative pixel values, suggesting a small oversubtraction
of the sky background. Although this has negligible effect on our
analysis of the bar and bulge region, we corrected the image by
measuring the background in the regions furthest from the galaxy centre
along the minor axis.
The original IRAC1 mosaic of \citet{Barmby2006} has a pixel scale of 0.863
arcsec/pixel and a size of $16300 \times 7073$ pixels ($3.89 \times
1.69$ degrees), using here for the analysis a resolution of 8.63\as/pixel. 
We fit ellipses to the 3.6\mum isophotes using the \texttt{ellipse}
task in the \textsc{stsdas} package in \textsc{iraf}; this task is based
on the algorithm of \citet{Jedrzejewski1987}. 
The algorithm fits ellipses to isophotes by minimizing the deviations in intensity around a given
ellipse; for an ellipse with a given semi-major axis $a$, this is done
by iteratively adjusting the ellipse centre (pixel position in $x$ and
$y$), orientation (position angle ${\rm PA}$), and ellipticity ($1 - b/a$ or $\epsilon$ , where
$b$ is the semi-minor axis of the ellipse). 
From \texttt{ellipse} we obtained the azimuthally average (AZAV) intensity ($I$) profile.
All magnitudes are in the Vega system, and we use the 3.6\mum absolute solar magnitude 
$M_{\odot}^{3.6}\e3.24$ \citep{Oh2008}.
We convert the intensity profile to surface-brightness profile (SB) using the 3.6\mum zero-point calibration
$280.9\pm4.1$ Jy \citep{Reach2005}.
For each fitted ellipse, the algorithm also expands the residual variations in intensity around
the ellipse in a Fourier series:
\begin{equation}
I(\theta) \; = \; I_{0} \sum_{m = 3}^{n} \left(\tilde{A}_m \sin m \theta +
\tilde{B}_{m} \cos m \theta \right),
\end{equation}
where $I_{0}$ is the mean intensity along the best-fit ellipse; the $m = 1$
and $m = 2$ components are automatically zero for the best-fit ellipse.
In fact, the actual computation divides the $m = 3$ and higher coefficients
by the local radial intensity gradient and the ellipse semi-major axis to
generate coefficients of radial deviation from a perfect ellipse:
\begin{equation}
\frac{\delta r (\theta)}{r} \; = \; I_{0} \sum_{m = 3}^{n} \left(A_m \sin m \theta +
B_{m} \cos m \theta \right),
\end{equation}
where $r = \sqrt{a b}$. The result is a set of higher-order terms
($A_{3}$, $B_{3}$, etc.) describing how the actual isophote deviates
from a perfect ellipse. The most interesting coefficients from our point
of view are $A_{4}$ and $B_{4}$. The most well-known of these is
$B_{4}$, which is $> 0$ when the isophotes are ``disky'' (lemon-shaped)
and $< 0$ when the isophotes have a ``boxy'' shape. The $A_{4}$ term is
non-zero when the boxy/disky shape is \textit{rotated} with respect to
the fitted ellipse; \citet{Erwin2013} (hereafter ED13) discuss how both
components are affected by the presence of B/P structures in
bars.\footnote{Note that in the ellipse-fit code of \citet{Bender1988},
the $\cos 4 \theta$ disky/boxy term is denoted by $a_{4}$, and is
related to $B_{4}$ by $a_{4}/a = \sqrt{b/a} B_{4}$.}
We used logarithmic spacing of the semi-major axes for fitted ellipses,
to combine high spatial resolution in the central regions and higher
$S/N$ at large radii. 
The \texttt{ellipse} profiles obtained for M31 are shown later in 
Figures \ref{fig:PAbarang} and \ref{fig:ellipse}.

\section{Method}
\label{sec:meth}
\subsection{Simulations}
\label{sec:meth:sim}
We want to explore a scenario where the bulge of M31 is a pure B/P bulge or a combination 
of a classical bulge and a B/P bulge, and in the last case, to constrain the properties of the classical 
component. That there are no complete analytical descriptions of B/P bulges that reproduce the vertical 
complexity of these structures, together with our interest in models which evolve in time and naturally 
develop the B/P bulge structure in equilibrium with a classical bulge, compels us to proceed with N-body 
simulations. In N-body models B/P bulges emerge from a disk that forms a bar which later buckles 
vertically, creating a peanut or boxy structure. This is a non-linear process which involves a 
redistribution of mass in the inner part of the initial disk, where the potentials of all components 
are involved \ie the disk, the initial bulge and the dark matter halo. 
It is not then possible to predict quantitatively the properties of a model after it evolves from 
its initial conditions. We therefore proceed to make a systematic exploration of the initial parameters 
with simulations where we change one parameter at the time. We separate here the discussion into 
the generation of the initial $N$-body models, and the generation of bars and B/P bulges in those models.

\subsubsection{$N$-body models: initial conditions}
\label{sec:meth:sim:init}
We use the software {\sc Nemo}, an array of several independent programs and tools to perform 
$N$-body experiments and analysis in stellar dynamics \citep{Teuben1995}.
To generate the particle models we select the program {\sc MaGalie} \citep{Boily2001} 
based on the method proposed by \citet{Hernquist1993} which solves the Jeans equations to generate 
galaxies close to dynamical equilibrium with several components, e.g. a bulge, a disk and a dark matter halo.
The code works in natural units, ergo the gravitational constant is set to $G\e1\uv^2\ud \um ^{-1}$, 
and the internal units for the variables are:  \um (mass), \ud (distance), \ut (time) and \uv (velocity). 
The exact values of the scaling factors for converting internal units into 
physical units vary between each simulation, as explained later in Section \ref{sec:meth:tech}. Typical values are 
$1\um\si5\times10^{10}\sm$, $1\ud\si2\kpc$, $1\uv\si300\kms$ and $1\ut\e1\ud\uv^{-1}\Gyr$ which is $\si7.0\Myr$.
This allows the scaling of the models to M31 by matching velocity and spatial scales independently. 

{\sc MaGalie} builds $N$-body dark matter haloes (DMHs) with different mass density profiles, a cored isothermal profile 
(used by AB06), or a Hernquist profile. Here we chose the latter,
as a convenient approximation to a Navarro-Frenk-White (NFW) DMH profile. It has convergent mass at 
large radii $r\!\rightarrow\!\infty$, and mimics the cuspy NFW profile in the inner parts 
of the halo as both density profiles behave as $\si r^{-1}$ within their 
respective scale radius \citep{Springel2005}. It is given by
\begin{equation}\label{eq:halo}
  \rho_{\rm halo} (r)\e\frac{M_{\rm H}\,(2\pi r_{\rm h}^3)^{-1}}{r/r_{\rm h}\left(1+r/r_{\rm h}\right)^3}
\end{equation} 
where $r$ is the radius, $M_{\rm H}$ is the mass of the halo at $r\!\rightarrow\!\infty$ and $r_{\rm h}$ 
is the scale length. We truncate the haloes at $r\e22\ud$, which defines the actual halo mass in the simulation
$M_{\rm h}\e M\left(r<22\ud\right)$. The density of the DMHs show some evolution in their
outer parts at 20\ud, but quickly stabilises within 100\ut, well before the bar formation, to its final shape.

The initial disk density profile is given by: 
\begin{equation}\label{eq:disk}
   \rho_{\rm disk} ({\rm R},z)= M_{\rm d}\,(4\pi h^2\,z_{\rm o})^{-1}\exp(-{\rm R}/h)\,{\rm sech}^2(z/z_{\rm o})
\end{equation}
where ${\rm R}$ is the cylindrical radius and $h$ is the initial radial scale length of the disk, 
which is fixed by {\sc MaGalie} to be $h\e1\ud$. 
The scale height of the disk is $\rm z_{\rm o}\e0.18\ud$ and the mass
of the disk is also fixed to $M_{\rm d}\e1.0\um$. As we are interested in the bulge we truncate the disk at $r\e10\ud$. 
The disks have an exponential radial dispersion profile. We choose a Toomre \citep{Toomre1964} value of 
$Q_{\rm T}{\left(R_{Q_{\rm T}}\right)}\e1.0$ measured at $R_{Q_{\rm T}}\e2.5\ud$ which avoids
axisymmetric instabilities, but allows non-axisymmetric instabilities to grow. 
We also modified {\sc MaGalie} to generate and test disks with an initial constant $Q_{\rm T}\e1.0$,  
as explained in Appendix \ref{sec:appA}. 

The initial bulges (ICB) are created also with a Hernquist density profile for which, as shown by \citet{Hernquist1990}, 
the projected surface-density profile agrees with a de Vaucouleurs profile (which is a S\'ersic 
profile with index $n\e4$), within $\approx35$ per cent for radii in the range $0.06\!\lesssim\! \rm R/R_{\rm e}\!\lesssim\!14.5$.
If integrated, this encloses $\approx94$ per cent of the total light. The density profile and parameters are defined here as:
\begin{equation}\label{eq:bulge}
   \rho_{\rm bulge} (r)=\frac{M_{\rm B}\,(2\pi r_{\rm b}^3)^{-1}}{r/r_{\rm b}\left(1+r/r_{\rm b}\right)^3}
\end{equation}
where $r_{\rm b}$ is the bulge scale length and $M_{\rm B}$ is the mass of the bulge at $r\!\rightarrow\!\infty$. 
We stop the particle sampling at $r\e2\ud$, which defines the actual ICB mass in the simulation 
$M_{\rm b}\e M\left(r<2\ud\right)$. 
During the evolution, the ICB density profile near the outer boundary evolves slightly, involving less than 4 per cent of 
the ICB particles.

\subsubsection{$N$-body models: bars \& Box/Peanut bulges}
\label{sec:meth:sim:bar}
Programs like {\sc MaGalie} can set up models of disk galaxies, but to study the possible coexistence 
of a B/P bulge with a classical bulge, we need to evolve the initial models to generate the required 
structures \citep{Athanassoula2005}. These N-body models generally form a bar that later 
goes through the buckling instability generating the B/P bulge or thick bar in the centre, which transitions 
to the thin bar further out that is aligned with the B/P bulge. Transient material trailing or leading 
the thin bar, like spiral arms attached to the thin bar ends, are not counted as part of the bar.
We reserve the term bar for the whole structure, that includes both the thin bar and the B/P bulge.

We want to generate and explore models with bars that show a wide range of boxy structure,
pattern speed, bar length and bar strength among others. Therefore, similarly to \citet{Bureau2005}, 
we choose different concentrations and masses for the DMHs, leading to models dominated by the mass 
of the disk (MD models), and models dominated by the halo (MH models). MH models usually develop 
long thin bars and their B/P bulges have a strong X-shape. MD bars are shorter, the thin bar can be 
very weak, and the B/P bulge has a more boxy shape. We generate and explore B/P bulge models that 
also include initial classical bulges, as explained in the next section. 

\subsubsection{$N$-body models: parameter space exploration}
\label{sec:meth:sim:paramspace}
We build two sets of models to make a systematic exploration of parameters with a total of 84 simulations.  
The first set (Set I) contains ICBs combined with B/P bulges and is built from initial models with bulge, disk and DMH 
components. Here we want to explore how the different ICBs affect the observational parameters. 
Therefore in this set we vary only the initial mass and size of the ICB component, choosing 12 
different masses $M_{\rm b}$ ranging from 0.05\um to 0.6\um with steps at every 0.05\um. For each 
chosen mass we also explore different sizes for the bulge using 6 values of $r_{\rm b}$ ranging from 
0.1\ud to 0.35\ud with steps at every 0.05\ud, ending with a total of 72 simulations for this set.
The DMH used in this set has a scale and mass of $r_{\rm h}\e20\ud$ and $M_{\rm h}\e8\um$.

The second set of models (Set II) contains pure B/P bulges and is built from just disk and DMH initial components. Here we 
try to generate buckled bars with different boxy structures by changing the concentration and the mass of the 
DMHs. Therefore we use 3 scale lengths $r_{\rm h}$ of 10\ud, 15\ud and 20\ud and for each $r_{\rm h}$ we 
explore 4 different masses $M_{\rm h}$: 6\um, 7\um, 8\um and 9\um, making a total of 12 simulations.

In addition to these 84 simulations, we have run 100 simulations with different sets of initial parameters,
such as disks with different $z_{\rm o}$,  others with cored isothermal DMHs, and others with an initial disk 
with initial constant $Q_{\rm T}\e1.0$, but found that our fiducial choices best reproduced the bulge of M31, 
and for conciseness we do not give further details of these simulations here.

Due to the difficulty of plotting the results of the analysis of 84 simulations, we proceed to show only 
three examples in the next sections, Model 1, Model 2 and Model 3, which belong to Set I and therefore they 
have the same initial DMH and initial disk. Models 1 to 3 have the same ICB scale length of $r_b\e0.15\ud$ 
and differ only in the mass of the ICBs, which are 0.25\um(Model 1), 0.05\um(Model 2) and 0.5\um(Model 3). 
We also show Model 0 which is a pure B/P bulge of Set II that has the same initial DMH and disk as the previous models. 
We will show later that Model 1 is our best model for M31's bulge of all the explored models.

\subsubsection{$N$-body models: time integration}
\label{sec:meth:sim:runs}
To evolve the initial models we used a program also 
contained in {\sc Nemo} called {\sc gyrfalcON} \citep{Dehnen2000}. Although this program is 
not parallelized, it is a fast, momentum-conserving tree-code. It uses the same internal units of 
{\sc MaGalie}. We choose a time step of $t_{\rm step}\e2^{-6}\ut$ 
$\approx\!1.56\!\times\!10^{-2}\ut$ and we evolve the initial models until 600\ut (\si4.65\Gyr), 
analysing all the models this standard time. We also analyse and compare some models 
at 500\ut, 700\ut, 800\ut and 1000\ut. We choose a tolerance parameter of $\theta_{\rm tol}\e0.5$.
For simplicity we use a constant softening parameter of $\epsilon\e0.05\ud$. 

The number of particles that we use in both sets for the disk, classical bulge (if present) and DMH are $N_{\rm b}\e10^6$, $N_{\rm d}\e10^6$ and 
$N_{\rm h}\e2\!\times\!10^6$ respectively, and therefore the respective particle masses for each component are different,
with values for Model 1 of $m_{\rm b}\e1.2\times10^4\sm$, $m_{\rm d}\e4.8\times10^4\sm$ and $m_{\rm h}\e1.9\times10^5\sm$.
To examine the effects of force resolution on our main results we have re-run simulations with 
new softening parameters using a 50 per cent smaller global $\epsilon$ and later a 50 per cent larger global $\epsilon$.
To test the effects of unequal particle masses we follow the prescription by \citet{McMillan2007}: 
the softening for each particle depends on its mass and on the condition of the maximum force  
($F\si m/\epsilon^2$) allowed between the particles, obtaining for Model 1 the softenings for the bulge, disk and halo 
$\epsilon_{\rm b}\e0.0125\ud\,(30\pc)$, $\epsilon_{\rm b}\e0.025\ud\,(60\pc)$ and $\epsilon_{\rm h}\e0.05\ud\,(120\pc)$.
In the resulting simulation with lower resolution we observed no significant variations, while in the simulations with higher resolution we observed 
that the bar formation is delayed by roughly \si100\ut, but the bar evolution, including the buckling, does not change significantly 
and therefore the age of the bar remains the same, and the results stay unchanged.

\subsection{Technique to obtain the best-matching model}
\label{sec:meth:tech}
In order to compare simulations with observations we also use \texttt{ellipse} on images 
generated from our simulations. From this we obtain semi-major axis $(R)$ profiles of 5 
parameters: the position angle (${\rm PA}$) of the fitted ellipse, the Fourier coefficients $A4$ and
$B4$ which measure the asymmetry and the boxiness, 
the ellipticity $\epsilon$ and the azimuthally average (AZAV) surface-density profile 
$\Sigma\left(R\right)$ for the simulation images.  
The position angle is measured anticlockwise with respect to the north celestial pole axis, 
where ${\rm PA}\e0\degree$.
We rely on the coefficient $B4$ to quantify the strength of the boxiness (negative $B4$) 
or diskiness (positive $B4$) of the isophotes.
As already mentioned in the Introduction, the S\'ersic index is useful to classify bulges 
as classical bulges ($n>2$) or pseudobulges ($n<2$),
and therefore we measured this parameter in M31 and in our models.
For this we fit a combination of a S\'ersic profile \citep{Sersic1968} and an exponential profile to $I\left(R\right)$ ,
obtained from M31, and also to $\Sigma\left(R\right)$ from the simulations:
\begin{equation}\label{eq:fit}
\begin{aligned}
   I(R) &= I_{\rm e}\exp\left(-b_{n}\left[\left(R/R_{\rm e}\right)^{1/n}-1\right]\right) + I_{d}\exp(-R/R_{\rm d}) \\
   \Sigma(R) &= \Sigma_{\rm e}\exp\left(-b_{n}\left[\left(R/R_{\rm e}\right)^{1/n}-1\right]\right) + \Sigma_{d}\exp(-R/R_{\rm d})   
\end{aligned}
\end{equation}
where $b_n\e 1.999\,n-0.3271$ \citep{Capaccioli1989},  $R_{\rm d}$ is the disk scale length, $n$ the S\'ersic Index, and $R_{\rm e}$ the effective radius
which corresponds to the half light (or half mass) radius of the S\'ersic profile. $I_{\rm e}$ and $\Sigma_{\rm e}$ correspond to the half light and half 
mass of the S\'ersic profile. Here we denote by $I_{\rm Sersic}\left(R\right)$ and $\Sigma_{\rm Sersic}\left(R\right)$ the component in Eq.\ref{eq:fit}
fitted by the S\'ersic profile. The fit of the parameters is performed with a non-linear least squares (NLLS) minimization method using a Levenberg-Marquardt algorithm,
where we explore a full suite of Monte Carlo NLLS realizations with a wide range of initial guesses over all fitted parameters, from which  
we estimate errors from the standard deviations of the solutions around the best values.

We convert the $\Sigma\left(R\right)$ of the models to surface-brightness dividing by a stellar mass-to-light ratio ($M/L$).
This is determined after the profiles of Eq.\ref{eq:fit} are fitted to M31 and the models, by scaling $\Sigma_{\rm Sersic}\left(R\right)$ 
of the models to the intensity of M31 measured at the effective radius of M31 ($R_{\rm e}^{\rm M31}$) which is $I_{\rm e}$,
\ie  $\Sigma_{\rm Sersic}\left(R_{\rm e}^{\rm M31}\right)\e(M/L)\,I_{\rm Sersic}\left(R_{\rm e}^{M31}\right)\e(M/L)\,I_{\rm e}$.

In order to find a best-matching model for M31 we define 6 observational parameters: 
(1) $\Delta {\rm PA}_{\rm max}$, corresponding to the difference between the maximum PA (${\rm PA}_{\rm max}$)
in the boxy region of the bulge and the PA of the disk ${\rm PA}_{\rm disk}$; 
(2) $R_{B_4\e0}$ that corresponds to the radius where $B_4\e0$ and the isophotes stop being boxy 
and start being disky; 
(3) $B_4^{\rm min}$ that quantifies the maximum boxiness of the boxy bulge; 
(4) $\epsilon_{R_{\rm e}}$ the ellipticity at $R_{\rm e}$; 
(5) the S\'ersic index $n$ and (6) the effective radius $R_{\rm e}$.
And finally, we consider an additional parameter, which is the velocity scaling \uv calculated from the 
line-of-sight maximum velocity dispersion (7) $\sigma_{\rm los}^{\rm max}$ measured in M31.

\begin{figure}
  \begin{center}
    \includegraphics[width=8.0cm]{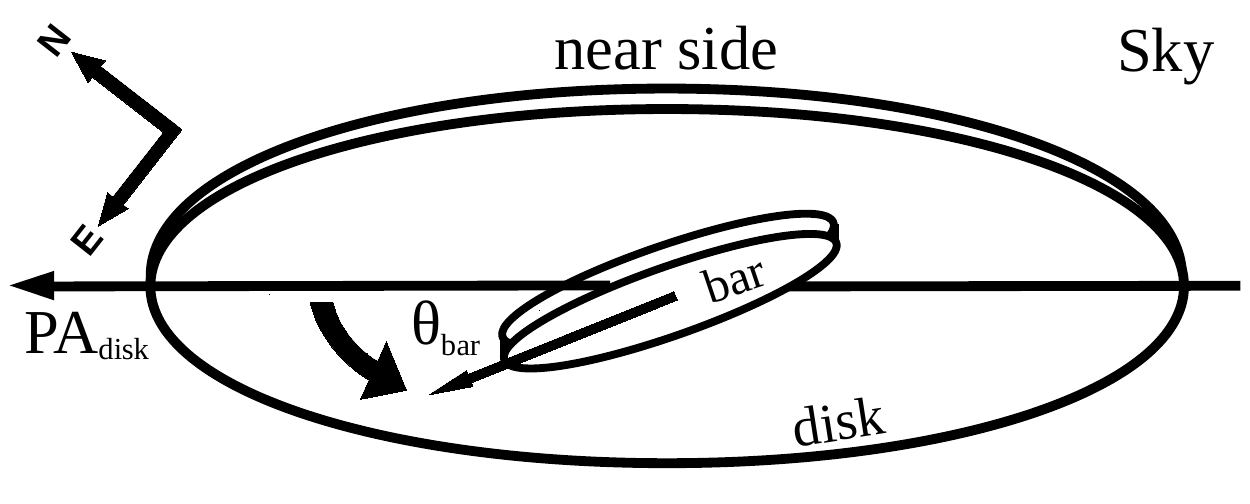}
    \vspace{-0.30cm}
    \caption{Schematic diagram of the orientation of the models. We project the models on the sky as M31, giving an inclination
    to the disk of $i\e77\degree$, locating the near side of the disk pointing to the north-west, and locating also
    the position angle of the projected major axis of the disk at ${\rm PA}_{\rm disk}\e38\degree$ anticlockwise from the north axis. 
    The bar angle $\theta_{\rm bar}$ is measured in the plane of the disk. 
    The straight arrow shows the major axis of the bar that is aligned with the projected disk major axis when 
    $\theta_{\rm bar}\e0\degree$. The angle $\theta_{\rm bar}$ increases anticlockwise, as shown by the curved 
    arrow, until for $\theta_{\rm bar}\e90\degree$ the bar is seen nearly end-on.}
    \label{fig:DiagramM31}
  \end{center}
\end{figure}

The procedure used to compare the observations with the models is the following:
\begin{enumerate}
\item 
We first project the models on the sky as M31, as shown in Fig.\ref{fig:DiagramM31}. 
For this we incline the disk to $i\e77\degree$ (where an edge-on disk is $i\e90\degree$, and a 
face-on disk is $i\e0\degree$). Then we rotate the projected model around the observer's 
line-of-sight axis until the projected disk major axis is aligned with the disk major axis of M31, 
leaving the position angle of the disk major axis like M31 at ${\rm PA}_{\rm disk}\e38\degree$ \citep{DeVaucouleurs1958},
and the near side of the disk in the upper part, pointing north-west like M31.
We specify the orientation of the model's bar by an angle $\theta_{\rm bar}$ in the plane of the disk, 
such that for $\theta_{\rm bar}\e0\degree$ 
the bar is side-on and its major axis is aligned with the projected disk major axis, and $\theta_{\rm bar}$ 
increases from the side of the disk major axis at ${\rm PA}_{\rm disk}\e38\degree$
in the direction away from the observer until for $\theta_{\rm bar}\e90\degree$ the bar major axis is almost aligned with the 
line-of-sight and is seen nearly end on.
Then we generate an image for each model, with a pixel size that slightly varies depending on the model, but with typical values of 5\as. 
\item We analyse the image of each model with \texttt{ellipse} and measure ${\rm PA}_{\rm max}$ in the boxy region.
${\rm PA}_{\rm max}$ is estimated as the error weighted mean of 5 ${\rm PA}$ measurements around the 
maximal PA value (with errors weights from the \texttt{ellipse} fitting), while for the ${\rm PA}_{\rm max}$ error
we use the error weighted standard deviation. This error is larger than the errors estimated by the 
\texttt{ellipse} fitting, and it takes into account the noise that we observe in the \texttt{ellipse} fitting profiles.
\item We repeat the previous step for each model, but with different $\theta_{\rm bar}$, ranging from 0\degree  until  74\degree,  
until the $\Delta {\rm PA}_{\rm max}$ of each model matches the observed value for M31 which is 
$\Delta {\rm PA}^{\rm M31}_{\rm max}\e13\degree\!\!.3\pm1\degree\!\!.2$, obtaining a 
best bar angle $\theta_{\rm bar}^{\rm best}$ for each model. This parameter is independent of the size scaling 
which only determines at what distance $\Delta {\rm PA}_{\rm max}$ is located.
To determine the error of the best bar angle we calculate
where $\theta_{\rm bar}$ matches the upper and the lower errors of $\Delta {\rm PA}^{\rm M31}_{\rm max}$,
from which we estimate the $\theta_{\rm bar}^{\rm best}$ error as $(\theta_{\rm bar}^{\rm up}-\theta_{\rm bar}^{\rm low})/2$ . 
As we show later in Fig.\ref{fig:PAbarang}, this error is larger than the effects of the 
noise in the ${\rm PA}$ profile on the bar angle estimation.
\item We use $\theta_{\rm bar}^{\rm best}$ for each model and we obtain the size scale \ud of each model by matching each 
$R_{B_4\e0}$ to the value for M31.
We show later in Section\ref{sec:res:morph} that the profile of the $B_4$ coefficient can successfully quantify the 
region where the boxy isophotes of M31 end ($B_4\e0$), and that our models exhibit a similar behaviour,
which makes this value ideal to our interest of restricting the size of the boxy region of M31's bulge 
and in our models.
\item We measure $B_4^{\rm min}$, $\epsilon_{R_{\rm e}}$, $n$ and $R_{\rm e}$ in each model.
\item We discard the models that do not match the selected 6 observational parameters of M31 
($\Delta {\rm PA}_{\rm max}$, $R_{B_4\e0}$, $B_4^{\rm min}$, $\epsilon_{R_{\rm e}}$, $n$ and $R_{\rm e}$), until we obtain 
a best model which simultaneously matches the parameters.
\item We obtain the velocity scale \uv of the best model by matching the maximum value of the line-of-sight dispersion profile
along the major axis of the of the model with the value measured from the bulge of M31 by 
\citet{Saglia2010}, \ie $\sigma_{\rm los} ^{\rm M31, max}\e\sigma_{\rm los} ^{\rm model, max}$.
\item We calculate dispersion and velocity profiles and maps for the best model and compare them with M31 observations.
\item We use the spatial and velocity scaling to obtain the mass scaling and calculate the mass profiles for the best model.
\end{enumerate}
At the end we obtain a model that matches the maximum position angle and the twist of the isophotes, with a boxy region
of similar extension and magnitude. And which contains a bulge with similar ellipticity, S\'ersic Index, and effective radius.
We discuss later the kinematic properties of the best model, which matches the central dispersion and the rotation observed in M31's bulge.

\section{Results}
\label{sec:res}
The results are organized in six subsections. 
Section \ref{sec:res:twist} explains how we match the isophotal twist of the bulge using the parameter $\Delta {\rm PA}_{\rm max}$, 
to determine the orientation of the bar in three-dimensional space ($\theta_{\rm bar}$).
In Section \ref{sec:res:morph} we show the results of our morphological analysis with 
\texttt{ellipse} on the observations and simulations, determining 3 parameters:
$B_4^{\rm min}$, $R_{B_4\e0}$, and $\epsilon$.
In Section \ref{sec:res:phot} we use the surface brightness profile of the observations 
and the surface mass profile to determine two parameters: the S\'ersic index and effective radius.
Section \ref{sec:res:par} compares the previous mentioned six parameters of the 72 models of Set I,
converging in a best matching model.
Section \ref{sec:res:bestmod} shows in more detail the properties of the best matching model,
showing its kinematics, mass profile and others.
Finally in Section \ref{sec:res:spur} we analyse the thin bar and the the spurs in the models and compare 
them with the observations.

\subsection{Morphology: bulge isophotal twist \& bar angle $\theta_{\rm bar}$}
\label{sec:res:twist}
\begin{figure*}
  \centering
  \includegraphics[width=17.5cm]{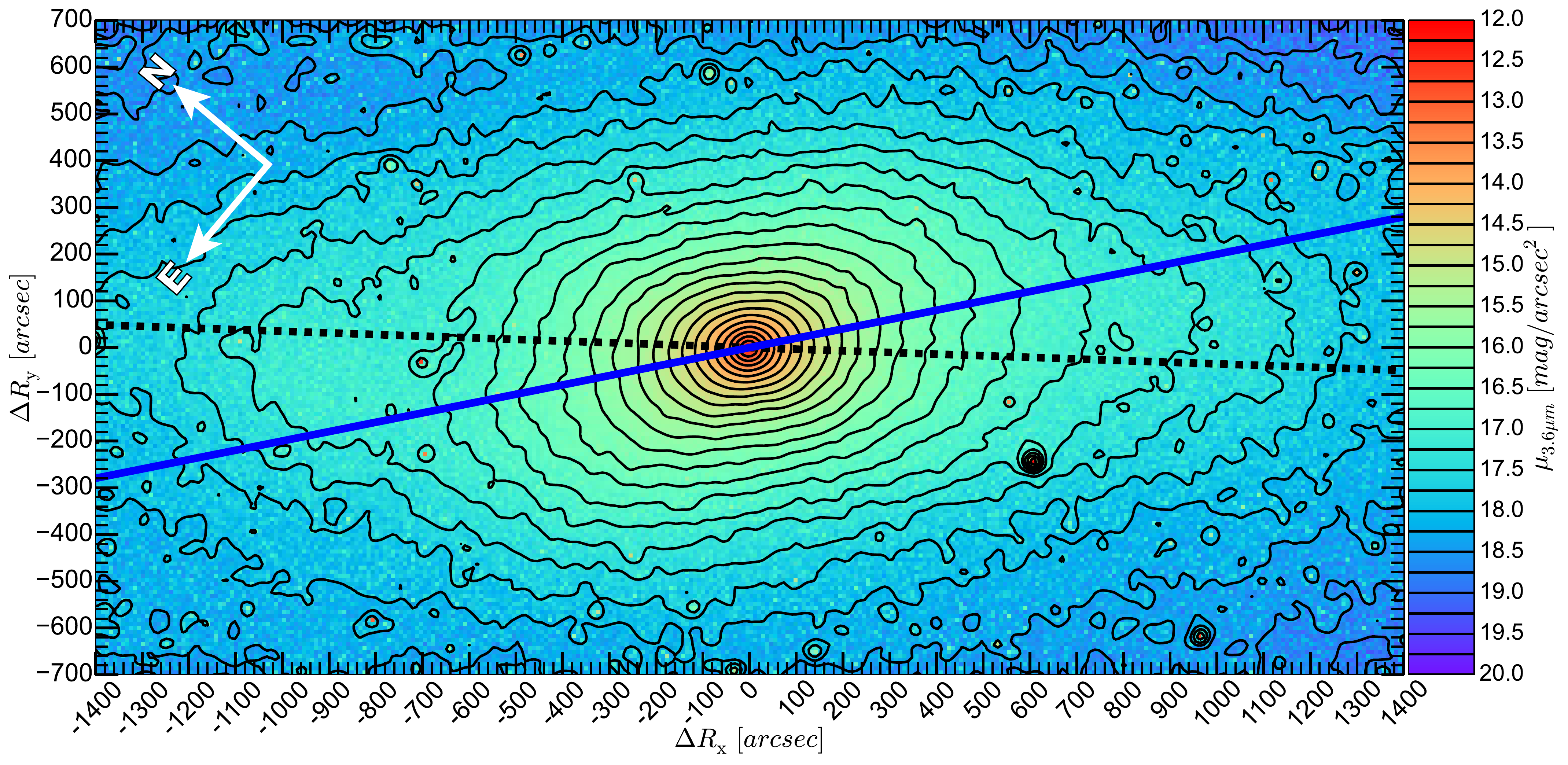}
  \includegraphics[width=17.5cm]{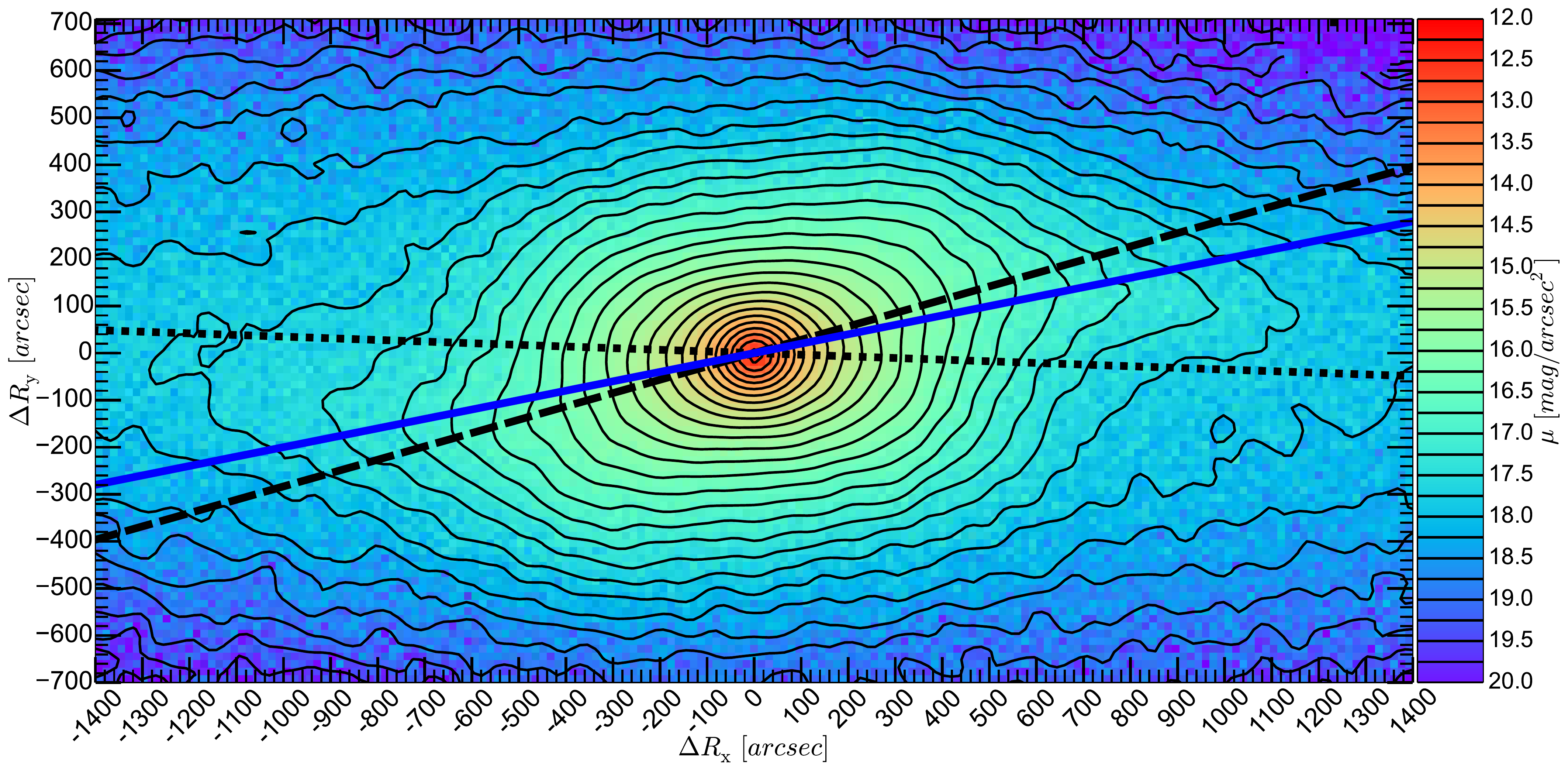}
    \vspace{-0.3cm}
    \caption{
    \textit{Top panel:}
    IRAC 3.6\mum image and isophotes of M31's central region. The position angle of 
    the horizontal axis $\Delta R_{\rm x}$ is ${\rm PA}_{\rm h}\e40\degree$. 
    The dotted line marks the position of the major axis of the disk (${\rm PA}_{\rm disk}\e38\degree$). 
    The blue solid line marks the maximum position angle determined by the fit of the ellipses, 
    reaching ${\rm PA}^{\rm M31}_{\rm max}\e51\degree\!\!.3\pm1\degree\!\!.2$, 
    and therefore a difference with the disk of 
    $\Delta {\rm PA}^{\rm M31}_{\rm max}\e13\degree\!\!.3\pm1\degree\!\!.2$, quantifying
    the clear isophotal twist in the central boxy region of the bulge. 
    The near side of the disk is located in the upper part of the panel \citep{Walterbos1988}.
    \textit{Bottom panel:}
   Image of the central region of the isophotes of Model 1 at 4.65\Gyr (600\ut). 
   The mass is converted to luminosity dividing by a stellar mass-to-light ratio of $M/L\e0.813\sm\slu^{-1}$.  
   We reproduce the central twist in the boxy region of the bulge, using the best bar angle 
   of $\theta_{\rm bar}^{\rm best}\e54\degree\!\!.7\!\pm3\degree\!\!.8$. The blue line marks the maximum 
   ${\rm PA}$ of the model which matches ${\rm PA}^{\rm M31}_{\rm max}$. The dashed 
   line marks the projected bar major axis for the best bar angle 
   ${\rm PA}_{\rm bar}\e55\degree\!\!.7\pm2\degree\!\!.5$.
   The near side of the disk is also located in the upper part of the panel. }
    \label{fig:SBmapM31}
\end{figure*}

\begin{figure}
  \begin{center}
  \includegraphics[width=8.5cm]{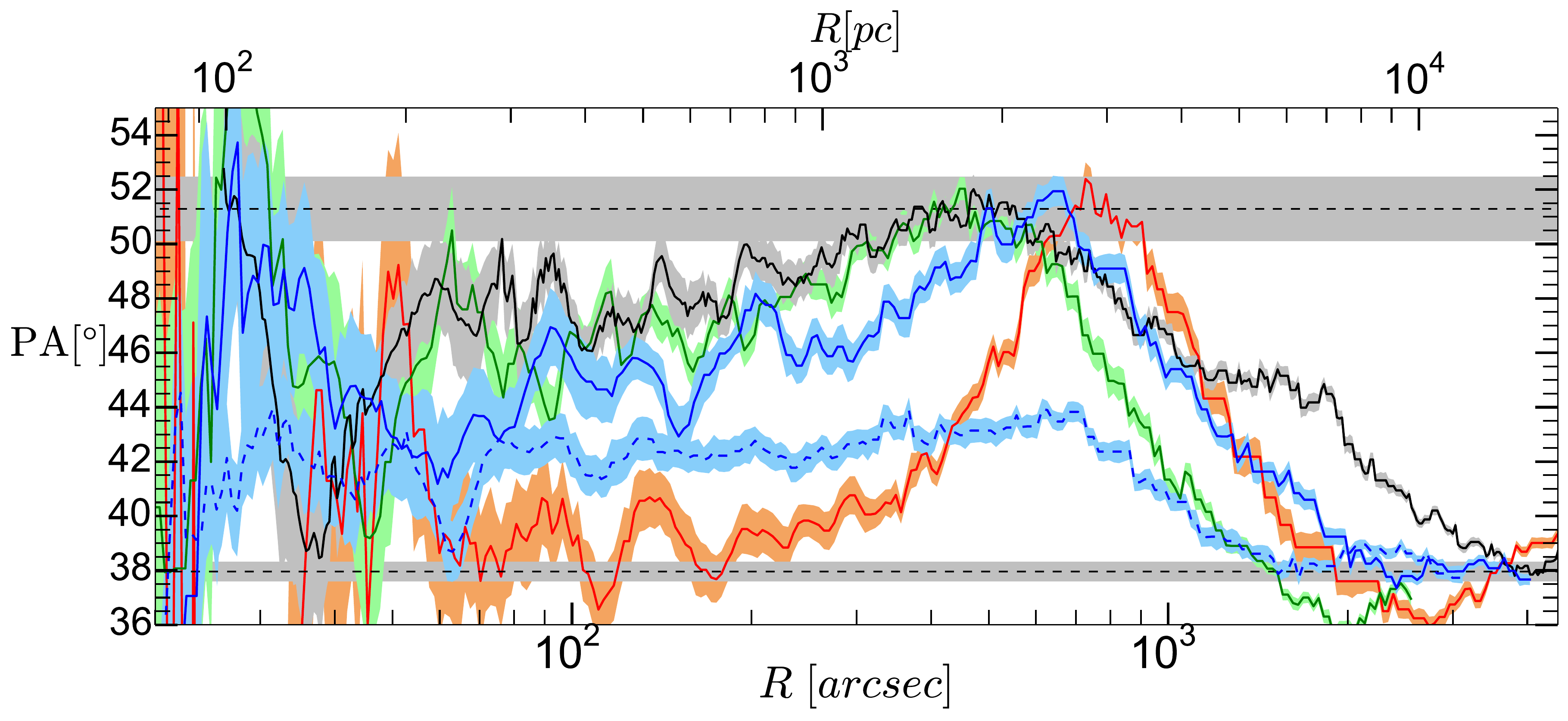}
  \includegraphics[width=8.6cm]{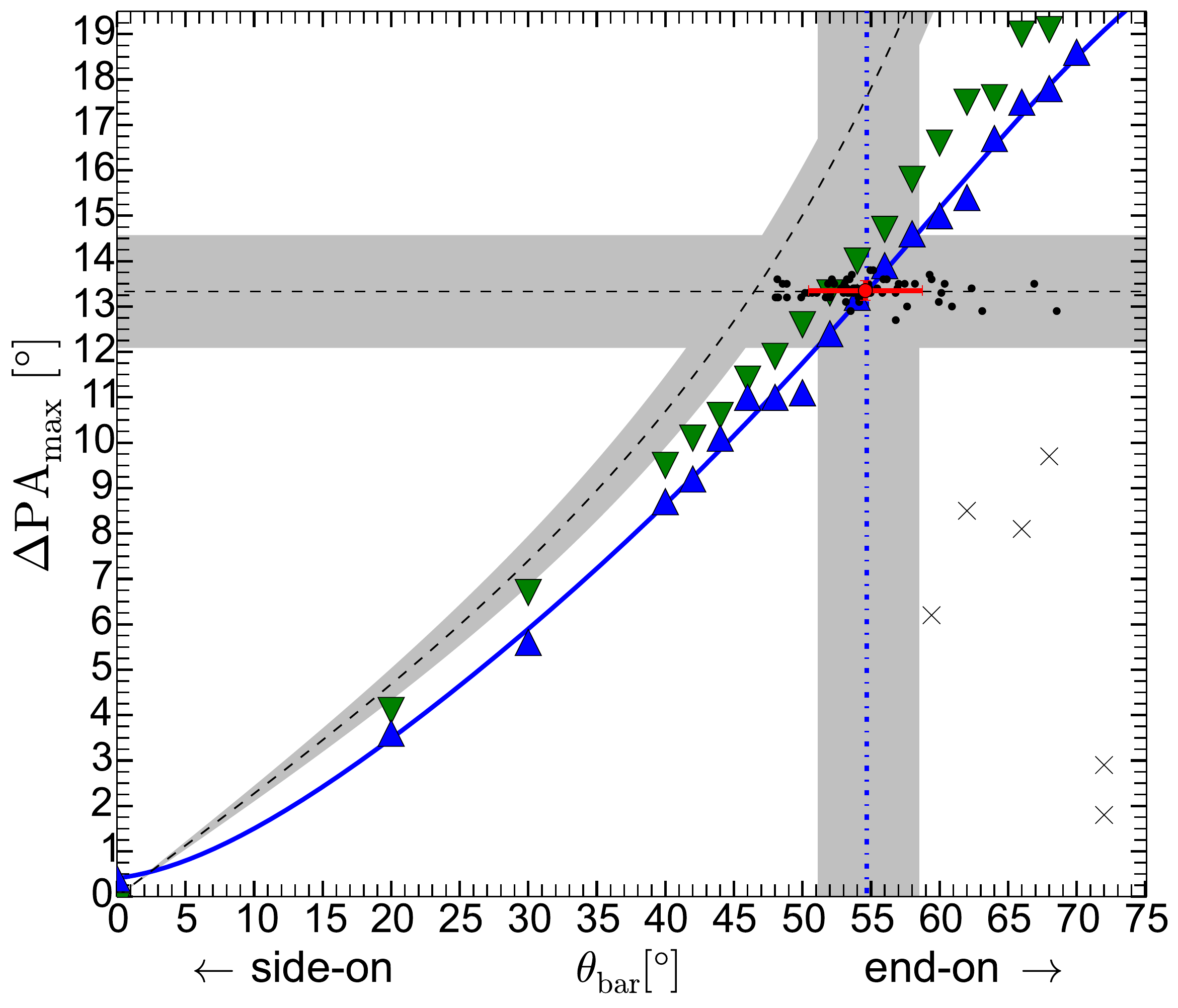}  
  \vspace{-0.6cm}
  \caption{
    \textit{Top panel:}
    The ${\rm PA}$ projected semi major-axis profiles determined with 
    \texttt{ellipse} for M31 (black), Model 2 (600\ut) (green) with 
    $\theta_{\rm bar}\e52\degree$ and Model 3 (600\ut) (red) with 
    $\theta_{\rm bar}\e66\degree\!\!.9$, with errors in shaded regions.     
    We show two profiles for Model 1 (600\ut), one where $\theta_{\rm bar}\e30\degree$ (blue dashed-line), and  
    another where $\theta_{\rm bar}\e54\degree\!\!.7$ (solid blue). 
    The horizontal upper and lower dashed black lines marks the ${\rm PA}$ peak of M31 
    ${\rm PA}^{\rm M31}_{\rm max}\e51\degree\!\!.3\!\pm\!1\degree\!\!.2$,
    and the ${\rm PA}$ of the disk of ${\rm PA}_{\rm disk}\e38\degree$.
    \textit{Bottom panel:}
    $\Delta {\rm PA}_{\rm max}$ versus the bar angle $\theta_{\rm bar}$. 
    When $\theta_{\rm bar}\e0\degree$ the bar is seen side-on, and 
    when $\theta_{\rm bar}\e90\degree$ it is end-on.
    The horizontal dashed line marks $\Delta {\rm PA}^{\rm M31}_{\rm max}$.
    The dashed curve shows $\theta_{\rm proj}\left(\theta_{\rm bar}\right)$ 
    of Eq.\ref{eq:angle}, where the 
    grey area shows the value when $i\e77\degree\!\pm\!1\degree$.
    Model 1 $\Delta {\rm PA}_{\rm max}$ values are shown for different $\theta_{\rm bar}$ (blue up-side triangles).
    The blue curve is a polynomial fit to estimate the best $\theta_{\rm bar}$
    that matches $\Delta {\rm PA}^{\rm M31}_{\rm max}$ for Model 1. The blue vertical dash-dot line marks Model 1 
    best bar angle $\theta^{\rm best}_{\rm bar}\e54\degree\!\!.7\pm3\degree\!\!.8$, 
    where the grey area are the values estimated from the observational errors.
    The down-side green triangles correspond to Model 2.    
    Dots and crosses correspond to the 72 models of Set I. 
    Dots mark the required $\theta_{\rm bar}^{\rm best}$ for each model to match 
    $\Delta {\rm PA}^{\rm M31}_{\rm max}$. The red cross shows the mean and standard deviation 
    for these best bar angle values: $\langle\theta_{\rm bar}^{\rm best}\rangle \e 54\degree\!\!.5\!\pm \!4\degree\!\!.5$.
    Black crosses are models that cannot reach $\Delta {\rm PA}^{\rm M31}_{\rm max}$,
    due to their concentrated ICBs that generate isophotes with low $\epsilon$.}
    \label{fig:PAbarang}
  \end{center}
\end{figure}

The goal of this step is to match the twist of the isophotes in the boxy region of M31's bulge
relative to the disk, as shown in Fig.\ref{fig:SBmapM31} (top panel). 
In external galaxies and in $N$-body simulations the presence of a classical bulge, and 
the orientation of the thin and the thick (or B/P bulge) components of the bar determine the shape 
and twist of the isophotes of the central region (\citealt{Bettoni1994}, AB06, ED13). 
ED13 compare surface-density contours of $N$-body barred disk models with 
isophotes of observed barred disk galaxies using the \texttt{ellipse} fitting analysis and identify 
several substructures that we also find in our models.

We quantify the twist in M31 and in the models using the difference between the maximum position angle 
${\rm PA}_{\rm max}$ in the boxy region and the disk ${\rm PA}_{\rm disk}\e38\degree$, obtaining for M31 
$\Delta {\rm PA}^{\rm M31}_{\rm max}\e13\degree\!\!.3\pm1\degree\!\!.2$. This is close to the measurements of 
\citet{Beaton2007} using the \texttt{ellipse} analysis on the bands J, H and \K of 2MASS 6X data, obtaining 
$\Delta {\rm PA}^{\K}_{\rm max}\si10\degree$.

We measured the $\Delta {\rm PA}_{\rm max}$ of the iso-density contours in each of our models using 
different angles for the bar $\theta_{\rm bar}$. 
In Fig.\ref{fig:PAbarang} (top panel) we show the ${\rm PA}$ profile and the ${\rm PA}_{\rm max}$ for M31 
in the boxy region compared to two ${\rm PA}$ profiles of Model 1 (0.25\um) that differ due to the different angles 
used for the bar, i.e. $\theta_{\rm bar}\e30\degree$ (blue dash curve) and $\theta_{\rm bar}\e54\degree\!\!.7$
(blue solid curve). As shown there, this model requires an angle for the bar of $54\degree\!\!.7$ 
to match the $\Delta {\rm PA}_{\rm max}$ observed in M31. We also show Model 3 (0.5\um) with the more massive ICB  
which needs an angle of $\theta_{\rm bar}\e66\degree\!\!.9$ in order to match ${\rm PA}_{\rm max}$.

We show in Fig.\ref{fig:PAbarang} (lower panel) how the bar angle $\theta_{\rm bar}$ of a 1D-bar 
measured in the plane of the disk changes its projection into a plane in the sky, with an inclination angle 
for the disk of $i\e77\degree$, described by the equations: 
\begin{align}
\label{eq:angle}
&\theta_{\rm proj}=\arctan (\tan \left(\theta_{\rm bar}\right) \cos\left(i\right))\\
\label{eq:PAbar}
&{\rm PA}_{\rm bar} \!= \theta_{\rm proj}\left(\theta_{\rm bar}\right)+{\rm PA}_{\rm disk}
\end{align}
where $\theta_{\rm proj}$ is the projection of the bar angle $\theta_{\rm bar}$, and 
${\rm PA}_{\rm bar}$ is the \textit{true} position angle of the projected major axis of the bar, 
that includes the thin bar and the B/P bulge.
Therefore if we approximate M31's bulge as a 1D-bar structure, the required angle to match the 
photometric twist $\Delta {\rm PA}^{\rm M31}_{\rm max}\si \theta_{\rm proj}$ would be 
$\theta_{\rm bar}\si46\degree\!\!.4$. $N$-body bars and real galaxies are vertically extended 
and therefore, excluding extreme cases, the difference between the bulge maximum position 
angle and the position angle of the disk ($\Delta {\rm PA}_{\rm max}$) will usually reach lower values 
than the infinitesimally thin case ($\Delta {\rm PA}_{\rm max}\!\leq\!\Delta{\rm PA}_{\rm bar}$),
exactly as we show with our simulations in Fig. \ref{fig:PAbarang} (lower panel).
In the figure we plot $\Delta {\rm PA}_{\rm max}$ versus $\theta_{\rm bar}$ for the Models 1 and 2. 
Using a polynomial fit of order 7 to Model 1 we find that the angle for the bar for which the 
$\Delta {\rm PA}_{\rm max}$ matches the value observed in M31 is 
$\theta^{\rm best}_{\rm bar}\e54\degree\!\!.7\pm3\degree\!\!.8$,
where the errors are given from the fit using the observational errors, as explained in 
Section \ref{sec:meth:tech}. This angle generates a twist of the isophotes in the boxy region of
Model 1, as shown in Fig.\ref{fig:SBmapM31} (bottom panel).
The projected angle of a 1D-bar given by this best angle is 
$\theta_{\rm proj}\left(\theta_{\rm bar}^{\rm best}\right)\e17\degree\!\!.7\pm2\degree\!\!.5$,
with its error calculated from the average of the upper and lower error in $\theta^{\rm best}_{\rm bar}$.
This locates the ${\rm PA}$ of the thin bar (and the B/P bulge) at 
${\rm PA}_{\rm bar}\e \theta_{\rm proj}\left(\theta_{\rm bar}^{\rm best}\right)+{\rm PA}_{\rm disk}\e55\degree\!\!.7\pm2\degree\!\!.5$.
Applying the same procedure to Model 2 and Model 3 we recover the best angles 
$\theta^{\rm best}_{\rm bar}\e52\degree$ for Model 2 and 
$\theta^{\rm best}_{\rm bar}\e66\degree\!\!.9$ for Model 3.
Model 1 and 2 recover more similar values for $\theta^{\rm best}_{\rm bar}$. Model 3 needs a larger 
$\theta^{\rm best}_{\rm bar}$ to match $\Delta {\rm PA}_{\rm max}$, 
because this model has a more massive ICB that dominates the morphology in the central region,
and therefore the isophotes have lower $\epsilon$ and a less boxy shape and shows a low 
${\rm PA}$. Nonetheless, further out it can reach the observed $\Delta {\rm PA}_{\rm max}$.

Looking carefully at the $\rm PA$ profiles of the models in Fig.\ref{fig:PAbarang} (top panel)
clarifies why ${\rm PA}_{\rm max}$ is chosen to determine the bar angle $\theta_{\rm bar}$.
From all the $\rm PA$ values in a profile, ${\rm PA}_{\rm max}$ is the closest value to the estimation given
by Eq. \ref{eq:angle}   ($\Delta {\rm PA}_{\rm max}\left(\theta_{\rm bar}\right)\sim \theta_{\rm proj}\left(\theta_{\rm bar}\right)$).
The deviation of $\Delta {\rm PA}_{\rm max}\left(\theta_{\rm bar}\right)$ from  $\theta_{\rm proj}\left(\theta_{\rm bar}\right)$
is shown in Fig.\ref{fig:PAbarang} (bottom panel).
This behaviour is observed in all our models with bars and is the reason why we choose ${\rm PA}_{\rm max}$ 
as an indicator for the bar angle.
We successfully match the ${\rm PA}_{\rm max}$ of M31, although the exact radius of the model's
${\rm PA}_{\rm max}$ depends on the morphology and length of the thick and thin bar.
As a consequence of this choice, the isophotes of the model show a photometric twist slightly weaker 
than in M31 within the radius where ${\rm PA}_{\rm max}$ matches ${\rm PA}^{\rm M31}_{\rm max}$.
Later, in Section \ref{sec:res:sub:par:angle}, we show that our conclusions do not change when we increase
$\theta_{\rm bar}$ to produce a more pronounced isophotal twist in the inner part of the bulge region of Model 1.

We repeat this process for all the models, obtaining their respective $\theta^{\rm best}_{\rm bar}$. 
The mean and standard deviation of the 72 models of Set I is 
$\langle\theta_{\rm bar}^{\rm best}\rangle \e 54\degree\!\!.5\!\pm \!4\degree\!\!.5$
which shows that the angle does not change much from model to model. 
Furthermore, we see in Fig.\ref{fig:PAbarang} (lower panel) that $\theta_{\rm proj}$ of Eq.\ref{eq:angle}
is a good predictor as a lower limit for bar angles, because none of the 72 models reach values lower than 
$\theta^{\rm best}_{\rm bar}\e46\degree\!\!.4$ when they match $\Delta {\rm PA}_{\rm max}$.
There are some outliers which never match $\Delta {\rm PA}_{\rm max}$, reaching always lower values
due to the fact that their ICBs have too much mass and/or are too concentrated and their 
round isophotes dominate.

AB06 used four N-body models with different $\theta_{\rm bar}$ and compare the spurs generated by 
the projection of the thin bar of the models with the spur like features at $R\si1000\as$ along the
major axis of the disk in M31, and concluded that the angle for the bar is between 
$\theta_{\rm bar}\e20^{\rm o}$ and $30^{\rm o}$ depending on which model they used. Here instead 
we use the isophotal twist of the bulge, obtaining $\theta_{\rm bar}^{\rm best}\e 54\degree\!\!.7\pm3\degree\!\!.8$,
and we argue later in Section \ref{sec:res:spur} that structures at $R\si1000\as$ are not simply 
related to the spurs generated by the thin bar.

\subsection{Morphology: position angle, boxiness, ellipticity \& asymmetry}
\label{sec:res:morph}
\begin{figure}
  \begin{center}
    \includegraphics[width=8.7cm]{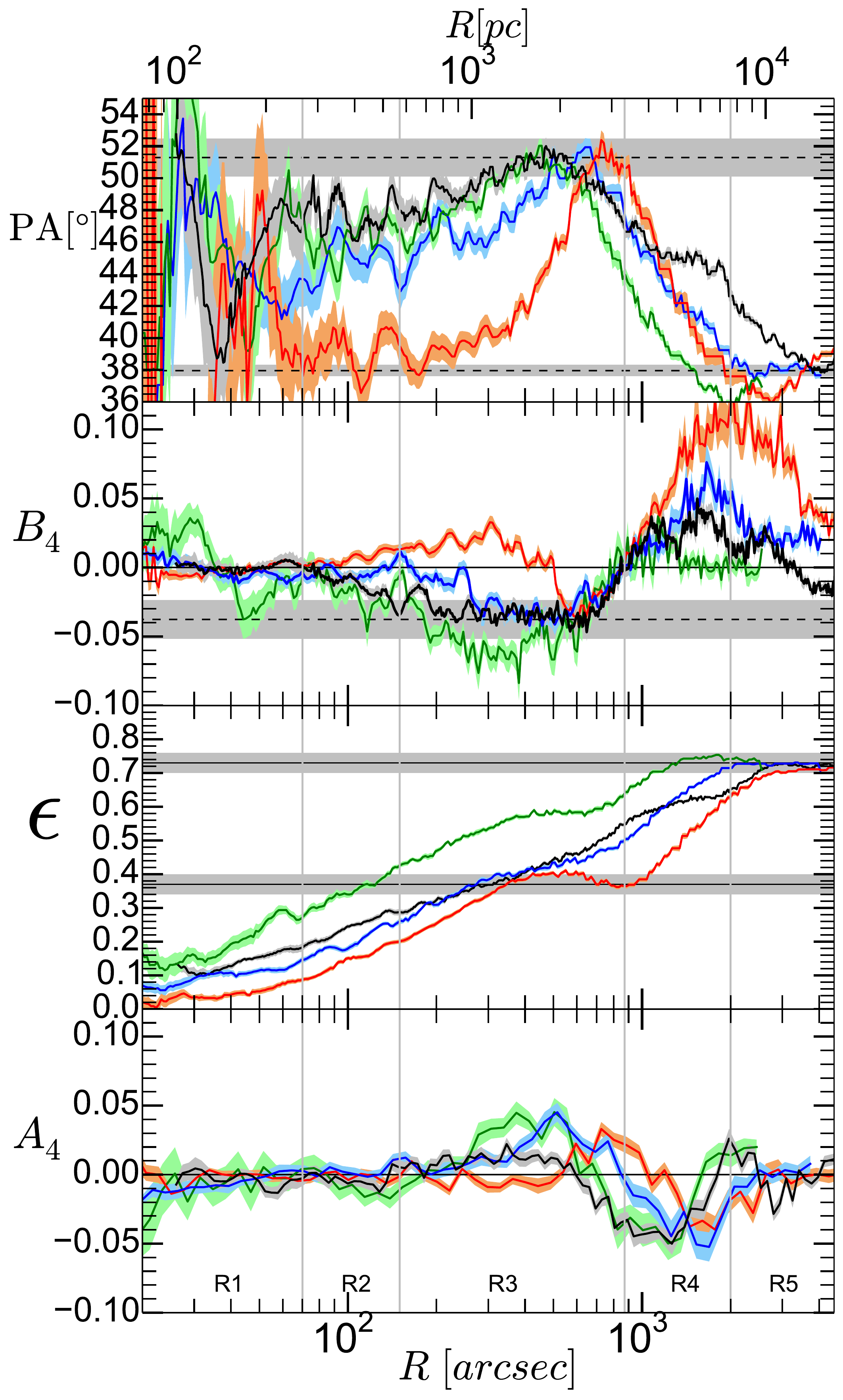}
    \vspace{-0.7cm}
    \caption{
    Parameters determined with \texttt{ellipse} for M31 (black),
    Model 1 (blue), Model 2 (green), and Model 3 (red) 
    including their errors (shaded areas), within the different regions (vertical lines).
    The profiles of the models correspond to a snapshot at 600\ut.
    \textit{Top panel}: ${\rm PA}$ profiles (same as Fig.\ref{fig:PAbarang})
    The upper and lower horizontal dashed black lines mark ${\rm PA}^{\rm M31}_{\rm max}$,
    and ${\rm PA}_{\rm disk}$.
    \textit{Second panel}: $B_4$ profiles. M31 reaches the maximum boxiness $B_4^{\rm min}\si-0.037$
    at $R\si600\as$ and its boxy region ends at $R_{B_4\e0}\e873\as$. The models
    are scaled in order to have the same $R_{B_4\e0}$, but their isophotes may have stronger or 
    weaker boxiness $B_4^{\rm min}$.
\textit{Third panel}: $\epsilon$ profiles. The upper and lower horizontal lines indicate 
$\epsilon$ for the disk and the bulge respectively by Co11 (see Section \ref{sec:res:phot}) where 
$\epsilon_{\rm bulge}\e0.37\!\pm\!0.03$ and $\epsilon_{\rm disk}\e0.73\!\pm\!0.03$. 
\textit{Bottom panel}: $A_4$ profiles. }
    \label{fig:ellipse}
  \end{center}
\end{figure}

In this section we compare the morphology of the isophotes of M31 with the N-body models.
For this we proceed in a similar way to ED13, with a morphological 
analysis using \texttt{ellipse} on the IRAC 3.6\mum image shown in Fig.\ref{fig:SBmapM31} 
and on images of our simulations, obtaining four parameters as a function of the 
major axis $R$ of the fitted ellipses: ${\rm PA}$, $B_4$, $\epsilon$ and $A_4$.
The resulting profiles for M31 and the models are shown in Fig.\ref{fig:ellipse}. 
We use these parameters to identify isophotal structures and to define five regions
(R1 ... R5) in M31, going from the inner part of the bulge to the disk.
Simultaneously, we plot in the figure the morphological profiles of M31 with three models, 
Model 1, Model 2 and Model 3 measured at 600\ut, to compare their 
properties within these regions.
These regions and the profiles are similar to the results 
obtained for M31's bulge in the bands J, H and K by \citet{Beaton2007}, 
and the morphological properties are also similar to what was found by ED13 
for their comparisons between B/P bulges of N-body simulations and other galaxies, using \texttt{ellipse}.

\begin{itemize}
 \item \textit{Region R1}: defined within $R\!\lesssim\!70\as$ where $A_4\si0$ and $B_4\si0$.
 M31 shows in this region isophotes with a low ellipticity $\epsilon\!\lesssim\!0.2$, symmetric, nearly-round isophotes, which
 makes the ${\rm PA}$ uncertain.
 Model 1 (0.25\um) shows similar values for $\epsilon$, $A_4$ and $B_4$ although the last two parameters indicate 
 slightly more symmetric isophotes than in M31. 
 Model 2 (0.05\um) has a higher $\epsilon$ and a more disky shape ($B_4>0$) 
 in the inner part ($R\si30\as$), but the boxy region starts sooner 
 ($R\si45\as$), and is also symmetric $A_4\si0$. This highlights the necessity
 of a more massive ICB component, like the one in Model 1, in order to obtain lower $\epsilon$ and a less boxy
 structure. But if the mass of the ICB is too high like in Model 3 (0.5\um) we observe a very low $\epsilon$ 
 due to the massive ICB, with large fluctuations of the ${\rm PA}$.

 Depending on the scaling parameter \ud of a simulation, region R1 and part of R2 could be within three softening lengths of the centre. 
 Our higher force resolution tests of Model 1 show only small variations of the surface-brightness profile in these regions (see Fig.\ref{fig:SBprofile})
 and we find that the isophotes and their morphological parameters change only slightly. 
 The small variation is in part because these are line-of-sight projected quantities  and therefore the differences are naturally smaller.
 
\item \textit{Region R2:} is the inner boxy region of M31's bulge, $70\as\!\!\lesssim\!R\!\lesssim\!\!150\as$,
defined by showing boxy isophotes with $B_4\si-0.02$, but still roughly symmetric with a low $A_4\si0$. 
Model 1 is less boxy than M31 in this region, but equally symmetric and with similar $\epsilon$. Model 2 is already 
more boxy than M31 and its $\epsilon$ is much larger than observed, reaching already $\epsilon\si0.45$.
The models differ again due to the presence of an ICB that 
dominates in this region, as shown later with the surface density and the mass profiles of Model 1 in 
Sections \ref{sec:res:phot} and \ref{sec:res:bestmod:mass}.
The ${\rm PA}$ profiles are better defined, reaching a value of $\si48\degree$ for M31,
$\si46\degree$ for Model 1 and $\si47\degree$ for Model 2, but still with some noise.
Model 3 still shows a very low ${\rm PA}$ and has more disky isophotes compared to M31, Model 1 and Model 2 .

\item \textit{Region R3:} is 
the outer boxy region of M31 with $150\as\!\!\lesssim\!R\!\lesssim\!873\as$, 
defined by the radius where the isophotes are boxy and start showing an asymmetry ($A_4>0$) and the radius
at $R\si873\as$ where $B_4$ changes from boxy (negative) to disky (positive), indicating the end
of the boxy isophotes and the transition to the inner region of the disk, we call this parameter
$R_{B_4\e0}\e873\as$. We also measure $R_{B_4\e0}$ in our simulations in internal units \ud and then we use 
the value of M31 to determine the scaling factor in our models. Therefore all the models end their boxy region
at the same place, although they do not necessarily do this in internal units. 
The maximum boxiness in M31 is $B_4\si-0.037$ located between $R\si573\as$ and $647\as$. 

Contrary to Region 2 this region shows an asymmetry of the isophotes in M31 given by a positive $A_4$ of $\si+0.1$, 
indicating that the ${\rm PA}$ is increasing. 
In our models we see the same features and behaviour of $A_4$ in this region, where two structures overlap which are 
the outer boxy bulge and the projected thin bar (that generates spurs). Further out the thin bar dilutes into the inner disk, 
sometimes with transient trailing or leading spiral structures. An increasing (decreasing) $A_4$ 
indicates that the isophotes are asymmetric and deviating anticlockwise (clockwise) from the major axis of fitted ellipse.

In this region we can see the twist of the isophotes in M31's bulge with respect to the isophotes of the disk, 
as shown in Fig.\ref{fig:SBmapM31}. This is reflected in the ${\rm PA}$ profile of the ellipses, which increase until a 
maximum of ${\rm PA}_{\rm max}\si51.3\degree\!\pm\!1.2\degree$. In the models we also see that the ${\rm PA}$ reaches a maximum 
in this region. 

The ellipticity of M31 at the effective radius ($R_{\rm e}^{\rm M31}\si365\as$) is $\si0.37$, which is in 
agreement with Co11, where $\epsilon_{\rm bulge}\e0.37\!\pm\!0.03$.
The $\epsilon$ profile of Model 1 and $\epsilon_{R_{\rm e}}$ agree quite well with the observations in this region. 
In Section \ref{sec:res:phot} we show for this model that the mass of the B/P bulge dominates over the ICB in this region, 
which has a strong impact on the shape of the isophotes. 
The profile of Model 2 shows high $\epsilon$, where $\epsilon_{R_{\rm e}}$ of this model almost doubles the value of M31. 
Model 3 has generally lower $\epsilon$ and more disky isophotal shape, and only at $R\si500\as$ starts showing a boxy shape.
At $\si400\as$ its ellipticity reaches the value of M31, but remains almost constant until the disk region.

\item \textit{Region R4:} 
This region shows the transition to the disk of M31 at $873\as\!\!\lesssim\!R\!\lesssim\! 2000\as$. 
The isophotes decrease their ${\rm PA}$ after reaching ${\rm PA}_{\rm max}$.
The ${\rm PA}$ shows a bump $\si45\degree$ between $\si1000\as$ and $\si2000\as$ which reveals a structure also visible in the 
isophotes in Fig.\ref{fig:SBmapM31}. 
The models shown here do not reproduce this feature, but we offer some possible explanations later in Section \ref{sec:res:spur}.
The ellipticity of M31 keeps rising, although near the bump it remains roughly constant, and again this is due to the same structure.
The ellipticity of Model 2 is higher than M31 and Model 1, reaching already the value of the outer disk $\epsilon\e0.73$.
Model 3 has much lower ellipticity than the observations, until $2000\as$.

\item \textit{Region R5:} the outer part of the disk ($R\!\gtrsim\!2000\as$), where the ${\rm PA}$ reaches $38\degree$,
aligning with the line of nodes of the disk.
The ellipticity reaches a maximum of $\epsilon\e0.73$ (also consistent with Co11, with $\epsilon_{\rm disk}\e0.73\!\pm\!0.03$). 
An infinitesimally thin disk would have an ellipticity of $\epsilon\e0.77$ using $i\e77\degree$. 
Our models have a vertically thick disk and reach $\epsilon\e0.73$, as in the observations.
\end{itemize}

\subsection{Photometry: M31's surface-brightness -- two bulge components?}
\label{sec:res:phot}

\begin{figure}
  \begin{center}
    \includegraphics[width=8.7cm]{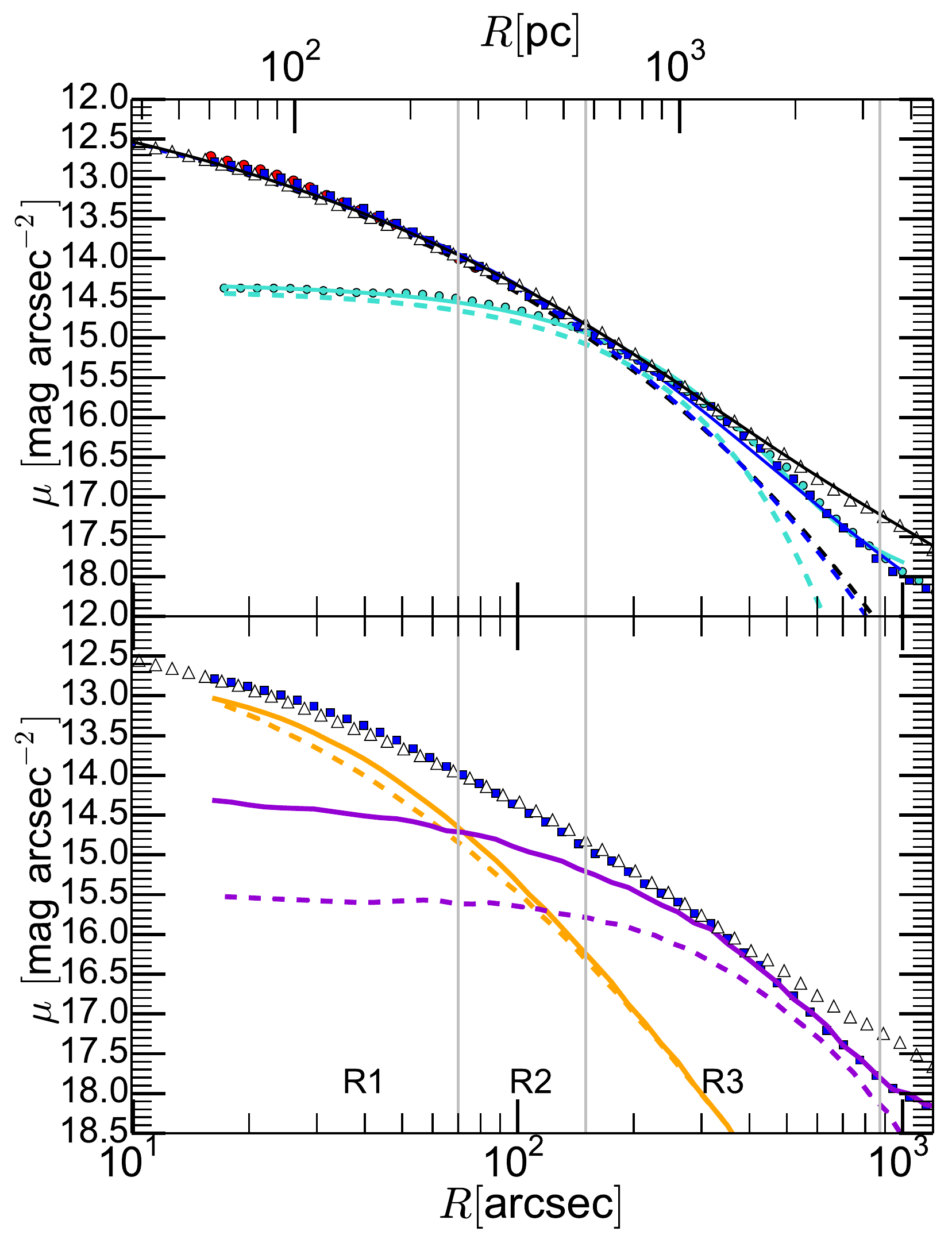}
    \vspace{-0.65cm}
    \caption{
    \textit{Top panel:}
    The surface-brightness (SB) profiles of M31 (white triangles), Model 1 (600\ut) (blue squares)
    and Model 0 (600\ut) (cyan circles), within the different regions (vertical lines). We include the 
    inner profile of a re-run of Model 1 with higher force resolution (red circles).
    The solid curves corresponds to the fit of a S\'ersic and an exponential profile, 
    and the dashed curve to the S\'ersic component alone.  
    The colours of the curves indicate if the fits belong to M31 (black), Model 1 (blue) or Model 0 (cyan). 
    A stellar mass-to-light ratio of $M/L\e0.813\sm\slu^{-1}$ is used in the models to convert $\Sigma\left(R\right)$ into SB.
    \textit{Bottom panel:} SB for M31 (white triangles) and Model 1 (blue squares), and 
    each component of the bulge of Model 1 plotted separately: the disk + B/P bulge (magenta solid curve) and the ICB (orange solid curve).
    We included also the bulge components of Model 1 at the initial time (dashed curves).}
    \label{fig:SBprofile}
  \end{center}
\end{figure}

In Fig.\ref{fig:SBprofile} we show the surface-brightness for M31's bulge region, for Model 0 
that is a pure B/P bulge (no ICB), and for Model 1 where we also show its components plotted 
separately (bottom panel). We also include in the figure our test of Model 1 
with a higher force resolution, finding a SB that only differs from the low resolution model 
by 0.5 per cent within 20\as. We determined the S\'ersic index and the effective radius for the 
models and for M31. We convert the surface density of Model 1 to SB dividing by a stellar 
mass-to-light ratio calculated as in Section \ref{sec:meth:tech},
$M/L\e0.813\sm\slu^{-1}$ ($\log_{10}(0.813)\e-0.09$), which agrees
with the range of values estimated by \citet{Meidt2014} for the IRAC 3.6\mum band, which goes from 
$\log_{10}(M/L_{3.6\mum})\approx-0.4$ to $\log_{10}(M/L_{3.6\mum})\approx0.04$ for a Chabrier IMF, 
depending on the metallicity and the age of the stellar population. The mass-to-light ratio shifts the profile 
of the models in the vertical axis of Fig.\ref{fig:SBprofile} and is an important value, 
but not critical at this step of the analysis, as it does not change the shape of the profile.

We see in the figure that the SB of Model 1 agrees quite well with M31 within Region 3 ($R\lesssim873\as$), 
where the B/P bulge dominates. Model 0 also shows a similar SB. Within Region 2 ($R\lesssim150\as$) M31 
keeps increasing its SB, while the pure B/P bulge, Model 0, does not. Model 1 matches the observed SB due 
to the contribution of its ICB. Within Region 1  ($R\lesssim70\as$) the SB of M31 keeps rising and it is 
matched by Model 1, where the ICB component now dominates with its cuspy SB over the 
B/P bulge component that shows a cored SB profile. It's interesting that Region 1 was defined with a 
morphological analysis of the isophotes of M31, and yet we see that this region is exactly where the SB 
of the classical bulge component dominates in Model 1. Model 0 shows also a cored SB and cannot reach the 
SB of M31, which highlights the importance of the ICB component and the mass concentration of this component.
We also show the SB of the two bulge components of Model 1 at the initial time, before the formation of the bar. 
The mass redistribution of the initial disk when its material forms the bar increases its SB in the central 
regions by $\si1\rm mag\,\as^{-2}$. The ICB almost does not change its SB despite the dynamical events 
like the bar formation and buckling, and the subsequent angular momentum transfer between components.

We also show in Fig.\ref{fig:SBprofile} the profiles that we fit to the SB of M31, Model 1 and Model 0.
From the fit to M31 we obtain the S\'ersic index $n\e2.6\pm0.8$, the effective radius $R_{e}\e1.4\pm0.5$\kpc, and 
the disk scale length of $R_{\rm d}\e5.7\pm2.1\kpc$.
$I_{\rm e}$ and $I_{\rm d}$ converted to surface-brightness are $\mu_{\rm e}\e16.5\pm0.4[{\rm mag\,\as^{-2}}]$ and 
$\mu_{\rm d}\e16.9\pm0.4[{\rm mag\,\as^{-2}}]$.
These values are consistent with the results of Co11, where they used several fitting methods 
on surface-brightness profiles generated with different fields or cuts of the image, and 
depending on the masking and the fitting method, finding $n$ values, varying 
from $1.66\pm0.03$ to $2.4\pm0.2$, $R_{\rm e}$ from $\si0.618\pm0.01\kpc$ to 
$\si1.1\pm0.1\kpc$ and a range of disk scale lengths of $R_{\rm d}\e4.75\pm0.01\kpc$ to $5.8\pm0.1\kpc$.

We find in our models that after the bar formation the surface-brightness profiles in the disk region 
shows a broken exponential profile, 
with two disk scale lengths. In the outer region ($4000\as<R$) we find a $R_{\rm d}$ 
that is similar to the initial one, while in the inner region 
($2000\as<R<4000\as$) we find a larger $R_{\rm d}$.
This is also present and explained by \citet{Debattista2006b} as part of the secular evolution due to the mass re-distribution 
of the disk after the bar formation.
Therefore, as we seek models of the bulge region of M31, we choose to leave the disk scale length
as a fixed quantity, using $R_{\rm d}\e5.7\pm2.1\kpc$, and
obtain for M31 a S\'ersic index $n\e2.59\pm0.16$ and an effective radius $R_{\rm e}^{\rm M31}\e1.4\pm0.2$\kpc. 
We then fit the models in the same way, leaving the parameters of the S\'ersic profile 
$\Sigma_{\rm Sersic}\left(R\right)$ and the central disk intensity ($\Sigma_{\rm d}$) free to vary. 
In this way we can compare the parameters of the S\'ersic component of observations with the simulations 
alone, which is the component that better quantifies the properties of the bulge's density profile.

We find that Model 0, without a ICB component,  has a S\'ersic index $n\e0.8\pm0.2$, lower than M31. On the other hand Model 1, 
with its ICB, behaves very similarly to M31, with $n\e2.37\pm0.47$ and $R_{\rm e}\e1.23\!\pm\!0.47\kpc$. 
We find that models of Set I that have higher $M_{\rm b}$ and smaller $r_{\rm b}$ than Model 1 tend to show $n$ higher than observed.
On the other hand, models of Set II, which are pure B/P bulges rarely show $n$ higher than \si1 \citep{Debattista2006b}, and therefore this set is
ruled out. In the outer parts, beyond Region 3, Model 1 starts to show lower densities than M31's disk. 
We find that our N-body disk models tend to decrease faster than the observed profile in M31.
This disagreement is important, but not critical as our interest in this paper is to find models 
for M31's bulge.

We also find that the S\'ersic index of a model can change with the orientation of the bar. 
When the bar is side-on ($\theta_{\rm bar}\e0\degree$) the
S\'ersic index can be higher, because the ICB has less material of the B/P bulge in the line of sight, emphasizing
its steep density profile. And when the bar is end-on ($\theta_{\rm bar}\e90\degree$) the S\'ersic index can be lower,
because more material of the B/P bulge is in the line of sight, which has roughly an exponential ($n\si1$) profile.
Furthermore, the S\'ersic index decreases also in time as the bar and B/P bulge grows incorporating more material,
hiding the ICB component (depending also on the orientation).

\subsection{Parameter space for the ICB}
\label{sec:res:par}
\subsubsection{Selection of the best model}
\label{sec:res:par:bestmod}
\begin{figure}
  \begin{center}
    \includegraphics[width=8.7cm]{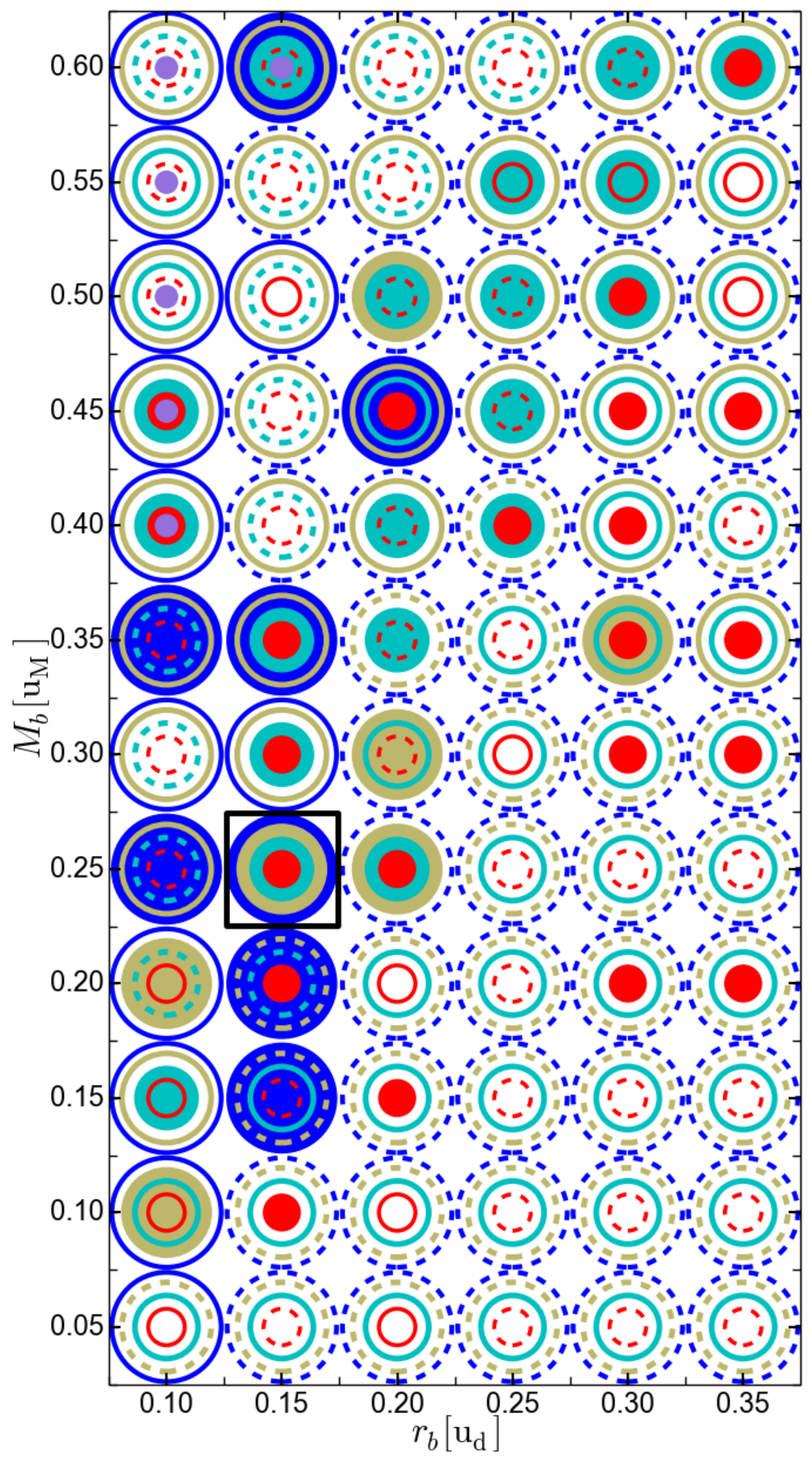}
    \vspace{-0.65cm}
    \caption{Model - M31 comparison in the parameter space of the ICB $M_{\rm b}$ versus size $r_{\rm b}$, 
    for the 72 models of Set I at 600\ut. 
    All models already match $R_{B_4\e0}$ and $\Delta {\rm PA}_{\rm max}$ 
    (except for 6 cases shown with a purple filled circle in the upper left corner).
    Blue, brown, magenta and red represent the parameters $n$, $B_4^{\rm min}$, $\epsilon_{R_{\rm e}}$ and $R_{\rm e}$.
    The parameters of the models that agree well with the parameters of M31 are shown with filled coloured circles, 
    and they have values that are within the range $-15\%\!<\!\Delta\!<\!15\%$.
    Solid thin circles mark models with values larger than the parameters in M31 by $15\%\!<\!\Delta$,
    while dashed circles mark values lower than M31 by $\Delta\!<\!\!-15\%$. 
    Model 1 matches simultaneously all parameters and is the best matching model (square).}
    \label{fig:MHRH}
  \end{center}
\end{figure}

As explained in Section \ref{sec:res:phot}, the pure B/P bulges do not show high mass concentrations in the centre and therefore their S\'ersic 
indexes are much lower than observed, all with values $n\lesssim1$, ruling out all models of Set II. 
Now we investigate the properties of the ICBs of Set I that can simultaneously 
reproduce all the morphological observables of M31's bulge. The most important initial parameters 
of the ICB are its mass $M_{\rm b}$ and its size $r_{\rm b}$. We show these two parameters in Fig.\ref{fig:MHRH} for 
the 72 models of Set I. The different colours used in the figure identify the four parameters: 
$B_4^{\rm min}$, $\epsilon_{R_{\rm e}}$, $R_{\rm e}$ and the S\'ersic index $n$. 
In order to compare the models and M31 we use the fractional difference between data and model (in percent): 
$\Delta\e 100\%\!\times\!(X_{\rm model}-X_{\rm M31})/X_{\rm M31}$ 
(where $X$ corresponds to the each of the four parameters).
If the circle is fully coloured with one colour it means that this parameter agrees to within $-15\%\!<\!\Delta\!<\!15\%$ of the value in M31.
Empty circles marked by a solid line (or dashed line) denotes that this parameter is larger than in M31 by $15\%\!<\!\Delta$ (or lower than in M31 by $\Delta\!<\!\!-15\%$).
The threshold of 15 per cent is chosen for two reasons: (i) typical errors in the parameters are between 10 and 20 per cent, 
and (ii) it is roughly at this value where is found a unique candidate that \textit{simultaneously} matches all the parameters.
By construction the models of Fig.\ref{fig:MHRH} already match $R_{B_4\e0}$ and $\Delta {\rm PA}_{\rm max}$ (except for 
a few models marked in the figure with purple circles that have ICBs that are too concentrated to match $\Delta {\rm PA}_{\rm max}$, 
as explained Section \ref{sec:res:twist}).
We highlight with a square in the figure the best model, Model 1, which simultaneously reproduces all the parameters.

The S\'ersic index gives information about the mass concentration of the models.
We explore a reasonable range of sizes and masses for the ICB, which give $n$
as low as 1.13 and as high as 3.23, with an average and standard deviation of $\langle n\rangle\e2.06\!\pm\!0.49$.
As expected the ICB with sizes smaller than $r_{\rm b}\!<\!0.15\ud$ are too concentrated, resulting in 
$n$ too high, with values between 2.52 and 3.23. 
\citet{Fisher2008} showed that pseudobulges and classical bulges can be distinguished by the threshold $n\si2$.
The pure B/P bulge models fail to reproduce M31 surface-brightness,
indicating that a classical bulge component is needed, but its mass contribution is limited in order to match $n$. 

The boxiness tends to be weaker in the upper right corner of the figure ($B_4\!>\!-0.037$), where the ICB are
more massive and concentrated, which makes the isophotes less boxy. Also, a compact bulge can 
help to stabilize the buckling instability, therefore making B/P bulge less boxy \citep{Sotnikova2005}. 
The inverse is also true, as shown by the models with less concentrated ICB and with more strongly boxy structure (or even X-shape) that are located 
in the bottom left corner of the diagram ($B_4\!<\!-0.037$).
The ellipticity behaves similarly to $B_4^{\rm min}$, where more compact ICB give round isophotes 
and therefore low $\epsilon_{R_{\rm e}}$.
We find that the variations of the parameters $B_4^{\rm min}$, $\epsilon_{R_{\rm e}}$ and $n$ vary more smoothly 
in the diagram, contrary to $R_{\rm e}$ which is more scattered.
The boxy structure forms in a non-linear process and its size can show a wide range of values of $R_{B_4\e0}$
in internal units (\ud), which in the case of Set I varies from 1.26\ud to 3.74\ud.
Since $R_{\rm e}$ is rescaled along with $R_{B_4\e0}$,
this results in a wide range for $R_{\rm e}$ as well.

\subsubsection{Best model parameters \& bar angle}
\label{sec:res:sub:par:angle}
In the previous section we analysed and compared the 72 models of Set I with parameters of M31, 
and at the end of the selection process only one model remains, Model 1. 
We found that the bar angle $\theta^{\rm best}_{\rm bar}\e54\degree\!\!.7$ best matches the 
$\Delta {\rm PA}_{\rm max}$ of M31.
As shown in Section \ref{sec:res:twist} matching $\Delta {\rm PA}_{\rm max}$ robustly fixes the bar angle.
This generates a twist of the isophotes in the outer part of the boxy region that matches M31.
In order to better match the twist in the inner part of the M31's boxy region however, 
Model 1 would require a larger angle. 
For this, we also tried $\theta_{\rm bar}\e60\degree$, and $65\degree$, resulting in a 
small variation of the parameters by less than $\vert\Delta\vert\!<\!15\%$.

Additionally, we tested $\theta_{\rm bar}\e35\degree$ and 45\degree,
finding that for 45\degree all parameters are in the range of $\vert\Delta\vert\!<\!15\%$.
For the most extreme case of 35\degree, the parameters $n$, $R_{\rm e}$ are still in the range of 
$\vert\Delta\vert\!<\!15\%$, while $\epsilon_{R_{\rm e}^{\rm M31}}$, $B_4^{\rm min}$ differ 
by 35 per cent from M31 values.
This is expected as at 35\degree the bar is orientated more side on, 
showing more elongated isophotes with its
B/P bulge appearing slightly less prominent. Of course, with 35\degree 
the twist of the isophotes in the bulge region is much weaker than in M31, 
which is reflected in the low ${\rm PA}$ profile shown for $\theta_{\rm bar}\e30\degree$
in Fig. \ref{fig:PAbarang} (top panel), and therefore we discard such low bar angle values.

\subsubsection{Best model parameters \& time evolution}
\label{sec:res:sub:par:time}
\begin{table}
\scriptsize 
\caption{Parameters for M31 and Model 1 at different times.}
\label{tab:param-time}
\begin{tabular}{lrrrrr}
\hline
Variable  								&  M31  &Model 1 	&		&			&	 \\  \hline
Time [\ut] (\Gyr)						& --	&500 (3.8)	&600 (4.6)	& 700 (5.4) 	&800 (6.2) \\
$\theta^{\rm best}_{\rm bar}$[\degree]	& --	&65.86		&54.7		&53.8		&54.5		\\
${\rm PA}_{\rm max}$ [\degree]				& 51.3 	&51.4 (0.8\%)	&51.3(0\%)	&51.7 (3.0\%)	&51.6(2.3\%)\\
$\Delta {\rm PA}_{\rm max}$ [\degree] 		& 13.3 	&13.4 (0.8\%)	& 13.3 (0\%)	&13.7 (3.0\%) 	&13.6 (2.3\%)\\
$B_4^{\rm min}$ 	$\times10^{-2}$		& $-$3.79	&$-$4.14 (9.2\%)	&$-$3.78 (0.3\%) &$-$3.98 (5.0\%)&$-$3.48 (8.2\%)\\
$\epsilon_{R_{\rm e}^{\rm M31}}$ 		& 0.37 	&0.31 (16.0\%)	&0.38 (5.2\%)	&0.38 (3.6\%)	&0.37 (0.1\%)\\
$n$									& 2.59 	&2.25 (12.9\%)	& 2.37 (8.3\%)	&2.44 (11.5\%)	&2.39 (12.1\%)\\
$R_{\rm e}$ [\kpc] 					& 1.40 	&1.31 (5.8\%)	&1.22 (12.3\%)	&1.03 (26.4\%)	&0.90 (36.0\%)\\
$\Omega_{\rm p}$ [$\uv\ud^{-1}$] (*)	& -- &  0.32 (41) 	& 0.29 (38) & 0.25 (32) & 0.23 (30)\\\hline
\end{tabular}
Notes: the percentages are the fractional difference between the observational parameters and the model parameters 
$\Delta\e 100\%\!\times\!(X_{\rm model}-X_{\rm M31})/X_{\rm M31}$ (where $X$ corresponds to the each parameter).
Units (*) in $\kms\kpc^{-1}$ using conversion scales of snapshot 600\ut.
\end{table}

We analysed the 72 models of Set I at the same internal time of 600\ut. 
We now analyse the temporal evolution of the parameters for Model 1.
As we explain later in Section \ref{sec:res:bestmod:kin:prof}, we convert $t\e600\ut$ to $t\e4.65\Gyr$, using a 
time scaling factor of $1\ut\e7.75\Myr$. This model formed the bar at $t\si200\ut\approx1.55\Gyr$, and 
therefore the bar has an age of $t_{\rm bar}^{\rm age}\approx400\ut\approx3.10\Gyr$ at the moment of 
our analysis. Following the same procedure we analyse Model 1 at 500\ut (3.87\Gyr), 700\ut (5.43\Gyr) and 
800\ut (6.20\Gyr), showing the results in Tab.\ref{tab:param-time} along with M31. 
We find that the snapshot at 500\ut shows slightly larger deviations in the parameters from the values 
observed in M31, compared to the snapshots at 600\ut, 700\ut 
and 800\ut, specially $\epsilon_{R_{\rm e}^{\rm M31}}$. There are two reasons: 
(i) at 500\ut the effects of the buckling instability are weak but still present in the bar, as shown later in 
Fig.\ref{fig:SBmap:snapshot}, 
and (ii) at 500\ut $\theta^{\rm best}_{\rm bar}$ is slightly increased, changing slightly the orientation of the bar.
But at later times the values only slowly change. Only the effective radius shows 
larger changes, but still within the observational errors estimated from our fit and by Co11.
This change is because the bar is slowly growing and in our method we scale the models to the size 
of the boxy region ($R_{B_4\e0}$). Therefore, a larger thick bar 
implies a smaller $R_{\rm e}$. 

\subsection{Best model properties}
\label{sec:res:bestmod}

\begin{figure}
  \begin{center}
    \includegraphics[width=8.8cm]{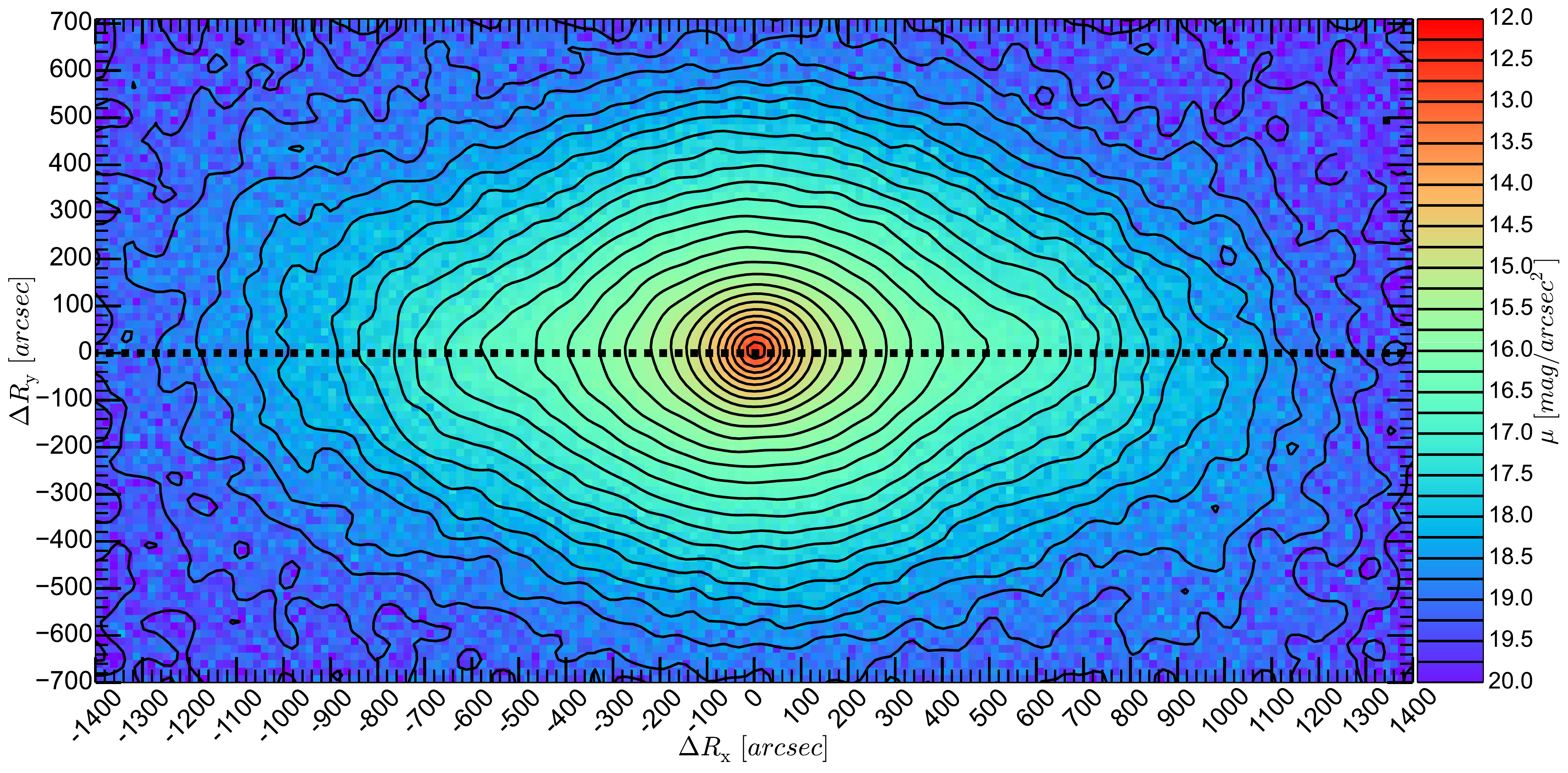}
    \includegraphics[width=8.8cm]{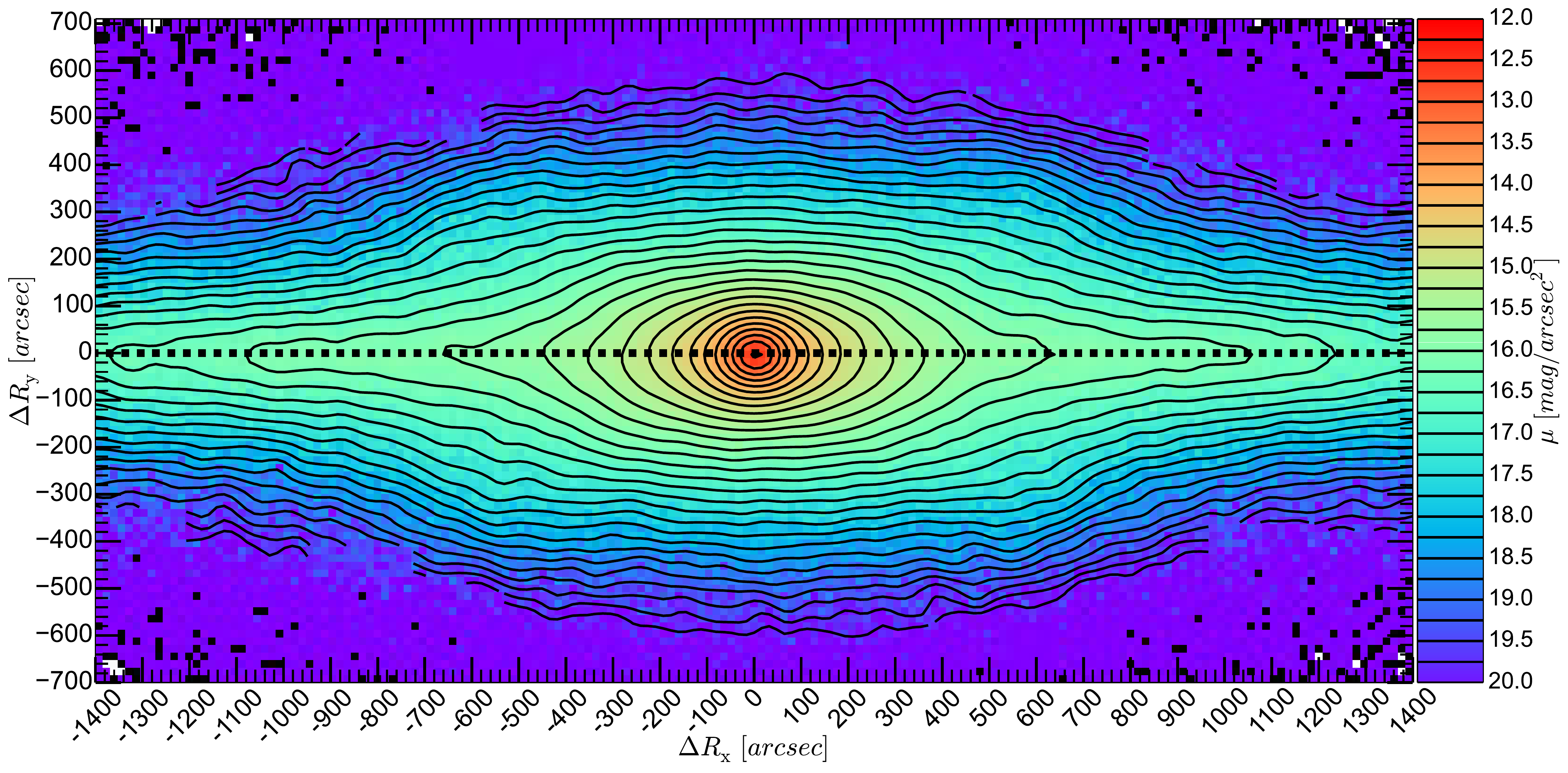}
    \includegraphics[width=8.8cm]{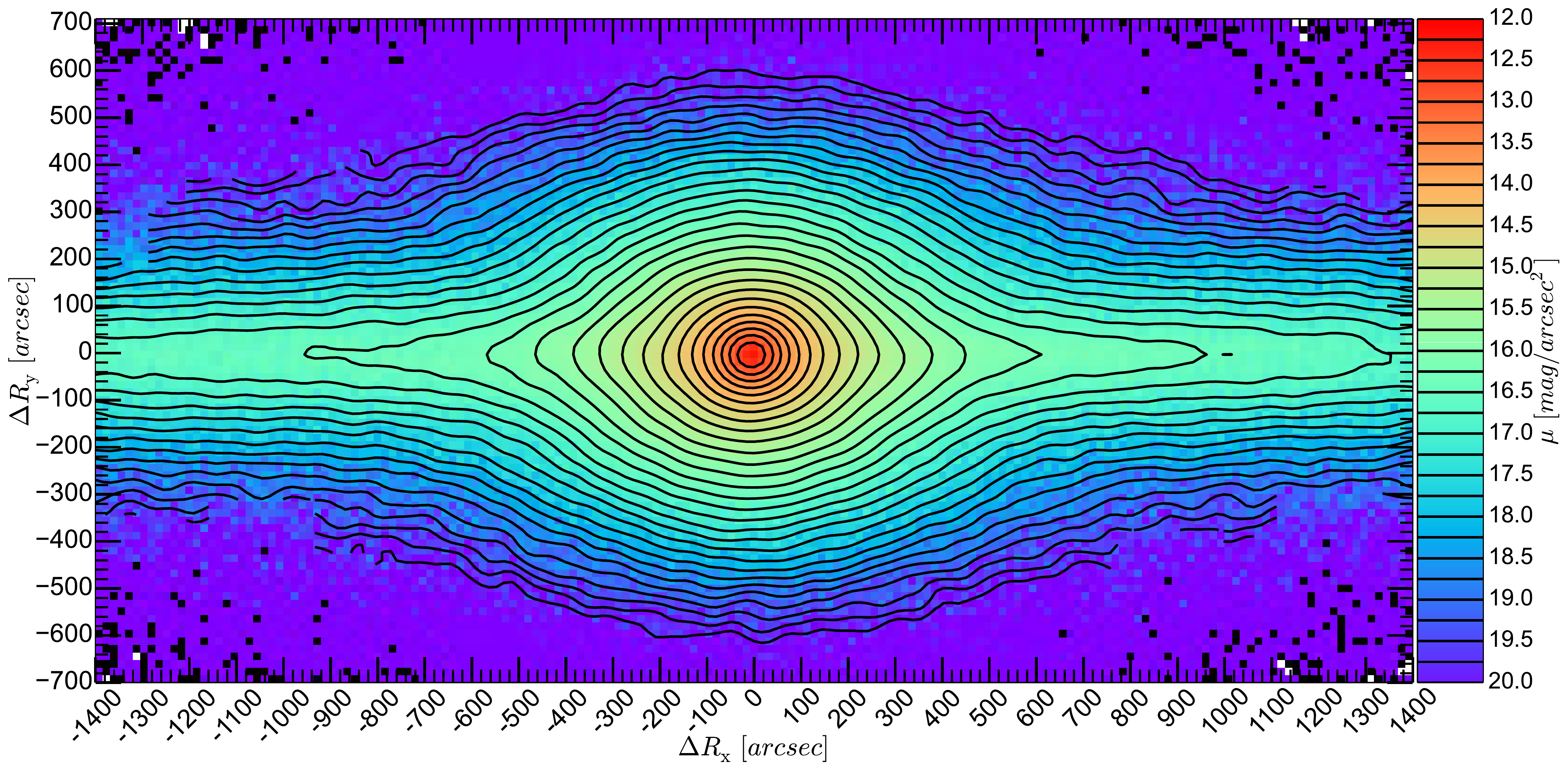}
    \vspace{-0.45cm}
    \caption{Model 1 SB maps at 4.65\Gyr (600\ut) observed with different orientations.
   \textit{Top panel}: the disk and bar with a face-on view ($i\e0\degree$, $\theta_{\rm bar}\e0\degree$), 
   where the thin bar extends between $\Delta R_{\rm x}\e-1300\as$ and $1300\as$. The dotted line shows
   the bar major axis.
   \textit{Middle panel}: the disk with a side-on view and the bar is seen edge-on ($i\e90\degree$, $\theta_{\rm bar}\e0\degree$),
   where it is possible to observe the boxy isophotes of the B/P bulge, extending from $\Delta R_{\rm x}\e-840\as$ to $840\as$ .
  \textit{Bottom panel}: the disk is side-on with the bar end-on ($i\e90\degree$, $\theta_{\rm bar}\e90\degree$).}
    \label{fig:SBmapMOD}
  \end{center}
\end{figure}

In Fig.\ref{fig:SBmapMOD} we show Model 1 at $t\e600\ut$ ($4.65\Gyr$) seen from different projections. 
The size scaling of the boxy zone using the parameter $R_{B_4\e0}$ gives a factor of $1\ud\e2.314\kpc$. 
The velocity scaling factor is determined by matching the maximum  $\sigma_{\rm los}^{\rm max}\e173.5\pm3.2 \kms$ 
of M31 along the slit of the photometric major axis of the bulge (${\rm PA}\e48\degree$).
The semi-major axis length of the thin bar measured in the plane of the disk from the centre is $r_{\rm bar}^{\rm thin}\e2.20\ud\e5.1\kpc$.
Later, in Section \ref{sec:res:spur:thinbar}, we show that the thin bar in M31 is shorter.
The B/P bulge is shorter than the thin bar, extending from the centre in the plane of the disk out to $r^{\rm B/P}\e3.2\kpc\,(840\as)$, 
and therefore it has a full length of $l^{\rm B/P}\e6.4\kpc\,(1680\as)$. Projecting the model 
on the sky like M31 ($i\e77\degree$, $\theta_{\rm bar}\e54\degree\!.7$), the B/P bulge radius is $R^{\rm B/P}\e1.9\kpc\,(510\as)$ 
and extends from end to end $L^{\rm B/P}\e3.8\kpc\,(1020\as)$.
We determined the B/P bulge 3D size using the prescription of ED13, from the average of the radius where 
$B_4\e0$ ($r_{B_4\e0}\e1192\as$), and the radius of $B_4^{\rm min}$ ($r_{B_4^{\rm min}}\e487\as$) 
measured along the disk major axis with $i\e60\degree$ and $\theta_{\rm bar}\e0\degree$ (bar seen side-on).

\subsubsection{Kinematics: $\sigma_{\rm los}$ and $\upsilon_{\rm los}$ profiles.}
\label{sec:res:bestmod:kin:prof}
\begin{figure}
  \begin{center}
    \includegraphics[width=8.6cm]{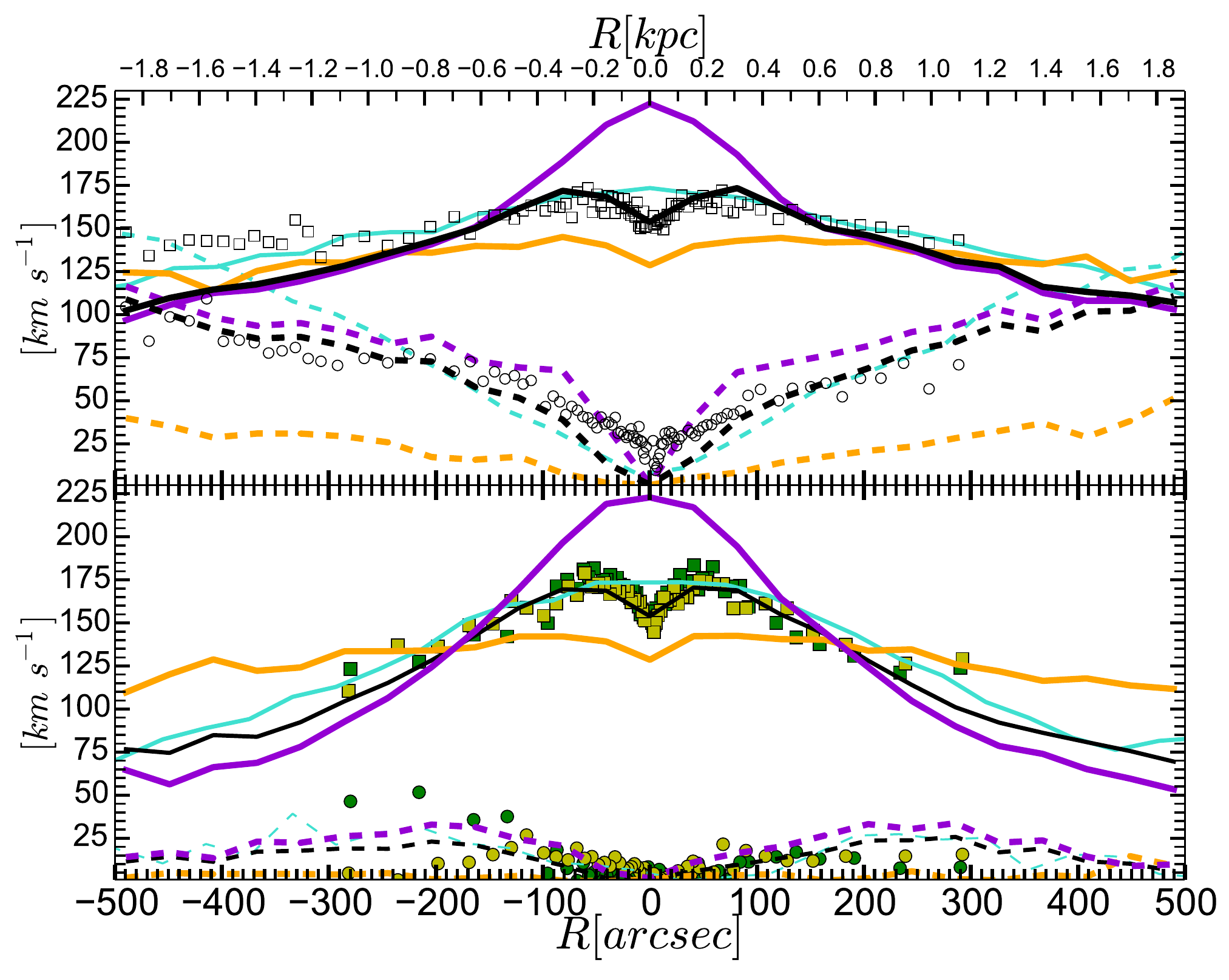}
    \vspace{-0.8cm}
    \caption{Line-of-sight kinematic profiles. We plot $|\upsilon_{\rm los}|$ (dashed curves)
    and $\sigma_{\rm los}$ (solid curves), for Model 0 (cyan) and for Model 1 (black) (both at 600\ut). 
    We also show the ICB component of Model 1 (orange) and the disk + B/P bulge component (purple).
    M31's  $\sigma_{\rm los}^{\rm fit}$ (circles) and $|\upsilon_{\rm los}^{\rm fit}|$ (squares)
    for different ${\rm PA}$.
    \textit{Top panel}: M31 values
    measured along the photometric major axis of the bulge at ${\rm PA}\e48\degree$ \citep{Saglia2010}.
    The kinematic profiles of Model 1 are calculated at the same ${\rm PA}$.
    Positive velocities ($\upsilon_{\rm los}\!>\!0\kms$) are located at the left side ($R\!<\!0\as$) and negative velocities
    at the right side.
    \textit{Bottom panel}: M31 values measured at the minor axis of the bulge, at ${\rm PA}\e138\degree$ (dark green),
    and values measured at ${\rm PA}\e108\degree$ (light green). Here the kinematic profiles for Model 1 
    are calculated at ${\rm PA}\e138\degree$.}
    \label{fig:kin:prof}
  \end{center}
\end{figure}

We compare the kinematic properties of Model 1 with the observations of \citet{Saglia2010}, who
determined the line-of-sight velocity distribution (LOSVD) along multiple slits located in the M31 bulge region.
They fit the LOSVD profiles with a Gauss-Hermite expansion following \citet{Bender1994}, obtaining the coefficients 
$\rm H_0\e1$, $\rm H_1\e0$, $\rm H_2\e0$, $\rm H_3$, $\rm H_4$, $\upsilon_{\rm los}^{\rm fit}$ and $\sigma_{\rm los}^{\rm fit}$, 
along slits at different ${\rm PA}$. 
Here we show three: one oriented at ${\rm PA}\e48\degree$, which is roughly aligned with the photometric major axis of the bulge
(or $+10\degree$ from the major axis of the disk);
another at ${\rm PA}\e138\degree$, which is aligned with the minor axis of the bulge, and one at ${\rm PA}\e108\degree$.
Once we project the models (using $\theta_{\rm bar}^{\rm best}\e54\degree\!\!.7$ and $i\e77\degree$), we use the same orientations of the 
slits of the observations to calculate for Model 1 and Model 0 the mass weighted line-of-sight velocity profile $\upsilon_{\rm los}$, 
and the mass weighted line-of-sight velocity dispersion profile $\sigma_{\rm los}$,
which are the first and the second kinematic moments, using a velocity scaling factor of 1\uv\e300.8\kms. 
The line-of-sight kinematic profiles for the ICB and the B/P bulge components of Model 1 are shown 
separately and combined in Fig.\ref{fig:kin:prof}, from which we see that they have quite different behaviour.  
The line-of-sight kinematic profiles of the B/P bulge are only slightly 
affected by the foreground and background disk material. In Section \ref{sec:res:bestmod:mass} 
we define a volume for the B/P bulge and use it to calculate the mass in the line-of-sight 
outside this volume \ie in the disk. We find along the slits at ${\rm PA}\e 48\degree$ and at 138\degree 
that on average only 13 per cent and 12 per cent of the stellar mass respectively is outside this volume, while within the 
effective radius 10 per cent of the stellar mass is outside this volume.

In the inner region ($R\!<\!150\as\si R_{\rm e}/2$), the B/P bulge component of Model 1 has a very high $\sigma_{\rm los}$ ,
reaching a peak of $\si220\kms$ in the centre. This is because the initial disk from which the B/P bulge emerges lives 
in the potential of the ICB component, and has a high amount of rotational kinetic energy. After the bar instability 
a large part of this energy in rotation is transformed into random motions (dispersion). 
In our test models with initial constant $Q_{\rm T}$ we obtained even higher $\sigma_{\rm los}$ in the centre for 
the B/P bulge component. The disk with higher \textit{initial} central dispersion (and therefore lower rotation) leads 
to a final central dispersion that is be lower than it would have been in the disk with a lower \textit{initial} central 
dispersion (and therefore higher rotation) see Appendix \ref{sec:appA}.
The buckling instability also increases the dispersion, but more slightly. 

The $\sigma_{\rm los}$ of the ICB component of Model 1 increases more slowly than the B/P bulge component, until $70\as$, where 
it decreases in an abrupt drop in the centre ($R<50\as$).
This is expected from the cuspy density profile. As we shown for the SB profiles, the ICB dominates in the centre, and 
therefore the combined profile also shows this feature. The $\sigma_{\rm los}$ measured in M31 also shows a drop in the centre. 
Model 0, which is a pure B/P bulge, does not show a drop in the centre, which is related to the cored $\Sigma$ profile previously shown in Fig.\ref{fig:SBprofile}.

Further out, at $R\!>\!R_{\rm e}$, the dispersion of the ICB component slowly decreases, while the B/P bulge component decreases faster.
The mass of the B/P bulge dominates in this region and therefore the combined profile follows the B/P bulge behaviour.

The velocity profiles behave differently for each component. 
We can see in the top panel of Fig.\ref{fig:kin:prof}, that although the ICB component had no rotation at the beginning, it shows 
some rotation due to the transfer of angular momentum from the bar, but it rotates much more slowly than the B/P bulge component.
In the central region $R\!<\!R_{\rm e}/2$ the rotation of the combined components is slightly lower than in M31,  
because the ICB dominates in this region, which could be corrected giving some initial rotation to the ICB.
At $R\!>\!R_{\rm e}$ the B/P bulge dominates and the combined bulge shows slightly higher $\upsilon_{\rm los}$ than in M31. 
This difference could be caused by some mass missing 
in the outer part of the bar and/or the disk, either dark matter, or stellar mass, as was also implied by the SB profile 
of Model 1, which drops slightly faster than that of M31 (Fig.\ref{fig:SBprofile}). 
A second reason could be the difference in the kinematic structure of the bar: Model 1 is slightly 
more supported by rotation than M31, while its dispersion is lower than in M31.
Fig.\ref{fig:kin:prof} (bottom panel) shows that M31 has some rotation along the minor axis of the bulge 
${\rm PA}\e138\degree$,  and shows even lower rotation in the slit at ${\rm PA}\e108\degree$. 
Our model also shows some rotation at 138\degree, and a lower rotation 
at 108\degree, which is related to the twist of the zero line-of-sight velocity curve, 
as shown in the velocity maps later.

\subsubsection{Kinematics: $\sigma_{\rm los}$, $\upsilon_{\rm los}$, $H_3$ maps \& the zero velocity line twist.}
\label{sec:res:bestmod:kin:maps}
\begin{figure}
  \begin{center}
    \includegraphics[width=8.8cm]{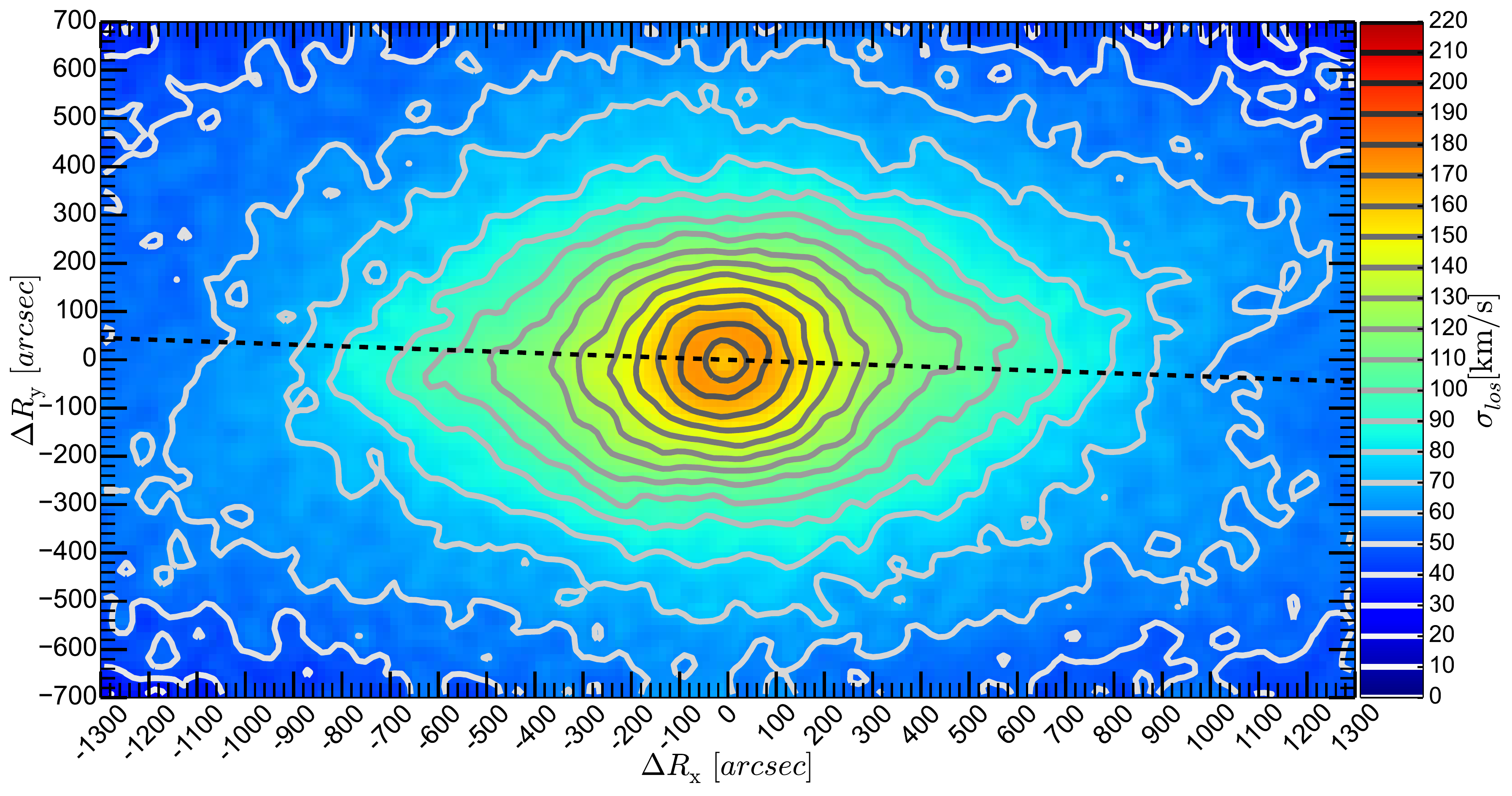}
    \includegraphics[width=8.8cm]{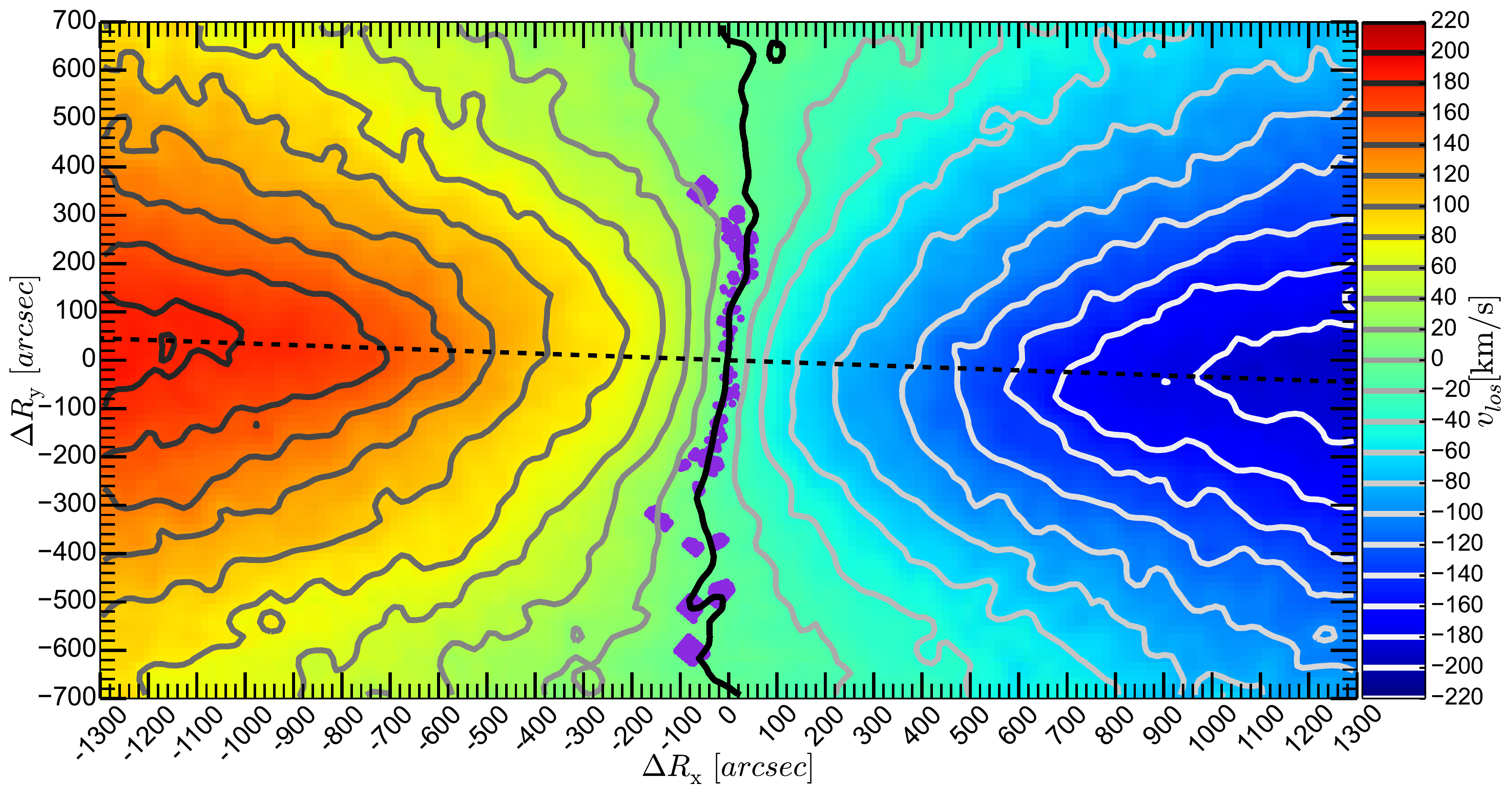}    
    \includegraphics[width=8.8cm]{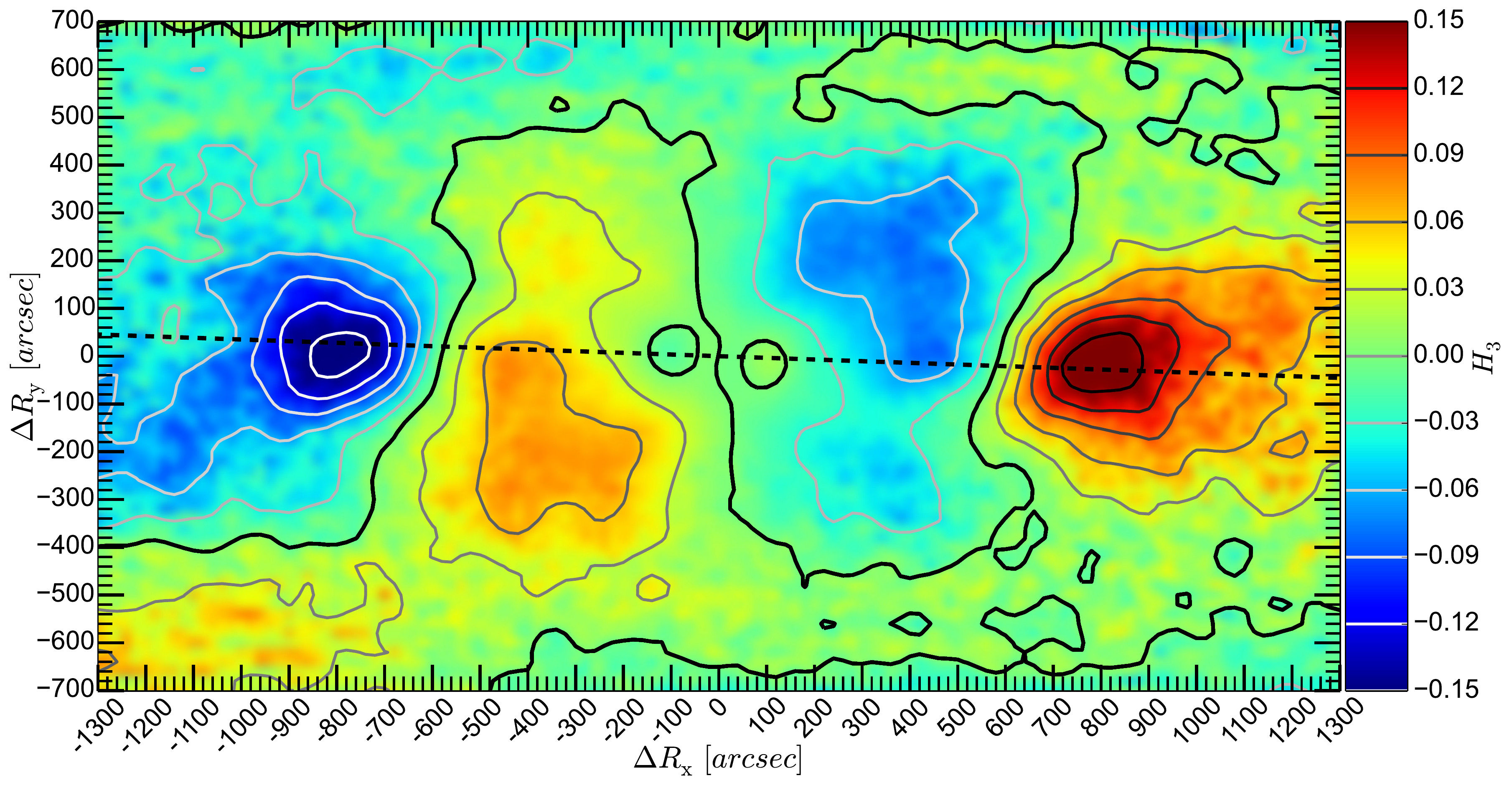}    
    \vspace{-0.65cm}
    \caption{Line-of-sight kinematic maps of Model 1 (600\ut), projected as M31 using $i\e77\degree$,  
    $\theta_{\rm bar}^{\rm best}\e54\degree\!\!.7$ and ${\rm PA}_{\rm disk}\e38\degree$ (dashed lines).
    The near side of the disk is in the upper part of the panels.   
    \textit{Top panel}: $\sigma_{\rm los}$ map and contours.
    \textit{Middle panel}: $\upsilon_{\rm los}$ map and contours. The thick black curve marks the 
    zero velocity line ($\upsilon_{\rm los}\e0\kms$) of the model. 
    The purple tracers correspond to the observed zero line-of-sight velocity determined by Opitsch et al. (in prep.).
    \textit{Bottom panel}: $H_3$ Gauss-Hermite coefficient map. The thick black curve marks the 
    zero $H_3$ values.}
    \label{fig:kinmap}
  \end{center}
\end{figure}

In Fig.\ref{fig:kinmap} we show $\sigma_{\rm los}$ and $\upsilon_{\rm los}$ maps of Model 1 in projection 
(using $\theta_{\rm bar}^{\rm best}\e54\degree\!\!.7$ and $i\e77\degree$), which reflect the same properties as 
the kinematic profiles of the last section.
Additionally, we add in Fig.\ref{fig:kinmap} the $H_3$ Gauss-Hermite coefficient map, 
because it allows to better distinguish the properties of the bar from the disk.

The dispersion map Fig.\ref{fig:kinmap} (top panel) also manifests a dispersion drop in the central region ($R<50\as$) 
due to the ICB component. Is also possible to 
observe that axis where the dispersion gradient is lower slightly differs from the photometric major axis of the disk.
The velocity field and contours exhibit nearly cylindrical rotation within a region of $R<200\as$. Trying different orientations for the bar ($\theta_{\rm bar}$)
we find that this model exhibits a stronger cylindrical rotation when the bar is end-on ($\theta_{\rm bar}\e90\degree$) and a weaker one 
when is side-on ($\theta_{\rm bar}\e0\degree$), due to the dynamics of a bar and to the presence of the ICB component.

Fig.\ref{fig:kinmap} (middle panel) shows the zero 
line-of-sight velocity contour in the central region at $\Delta R_{\rm x}\si0\as$ and 
going from $\Delta R_{\rm y}\e700\as$ to $-700\as$.  At these $\Delta R_{\rm y}$ 
the disk isophotes dominate, as shown in the photometry in Fig. \ref{fig:SBmapM31}.
Further inside the zero line-of-sight velocity contour shows twists at 
$\Delta R_{\rm y}\si300\as$ and at $-300\as$, due to the presence of the B/P bulge and its 
orientation, as shown also by isophotes in the photometry.
In the centre, within $\Delta R_{\rm y}\si100\as$ and $-100\as$
the twist becomes weaker due to the presence of the ICB component.
We also compare the zero line-of-sight velocity contour of the model 
with the observed velocities from Opitsch et al. (in preparation). 
Opitsch et al. observed the central 
region of M31 with the McDonald Observatory's 2.7-meter Harlan J. Smith 
Telescope using the VIRUS-W Spectrograph \citep{Fabricius2012}, 
which has a spectral range of 4850-5480 
Angstrom and a resolution of $R \approx 8700$ ($\sigma_{inst}\e15\kms$). They 
were able to completely cover the bulge region and also sample the disk out 
to one disk scale length (Co11) along six different directions, producing 
line-of-sight velocity and dispersion maps of the stars and the ionized 
gas. We plot here only the stellar velocities that are closer than 2\kms 
to the systemic velocity of $-300\kms$ \citep{DeVaucouleurs1991}.

We also include in Fig.\ref{fig:kinmap} (bottom panel) the predictions for the $H_3$ 
maps of Model 1. We find that $H_3$ and $\upsilon_{\rm los}$ generally anti-correlate in 
the disk region, \ie positive $\upsilon_{\rm los}$ are found where $H_3$ is negative, which 
is visible in the map along the major axis of the disk at $\pm700\as$ and beyond.
This can be seen as a consequence of the asymmetric drift.
It is also visible in the disk at different times, even in the inner region of the initial disk 
(within 700\as) before the bar forms. Later, after bar formation, we find that $H_3$ 
and $\upsilon_{\rm los}$ correlate within 700\as, which is where the bar is located 
\citep{Bureau2005}. This direct correlation between $H_3$ and $\upsilon_{\rm los}$ 
is also observed in M31 along the bulge photometric major axis exactly where the B/P bulge 
would be located \citep{Saglia2010}. In the left and the right sides of the $H_3$ map we
observe a sharp transition from negative $H_3$ to positive $H_3$ exactly 
where the B/P bulge ends and the disk starts, at $R_{\rm x}\si\pm600\as$ along the disk 
major axis. Also, $H_3$ is not symmetric with respect to the major axis of the disk, because 
the bar axis is oriented away from the disk axis, with a $\Delta {\rm PA}\si13\degree$.
In the B/P bulge region, $H_3$ reaches extremal values near the photometric major axis of 
the B/P bulge. The most extreme values in the entire plotted map are present in the disk region.
These qualities make $H_3$ a good parameter to estimate properties of the bar in M31.
Furthermore, we find in the central region ($R<100\as$) where the ICB dominates, that 
$H_3$ anti-correlates with $\upsilon_{\rm los}$ again, like in the disk, but much more weakly.

\subsubsection{Kinematics: circular velocity, pattern speed \& the Lindblad resonances.}
\label{sec:res:bestmod:kin:vcps}

\begin{figure}
  \begin{center}
    \includegraphics[width=8.2cm]{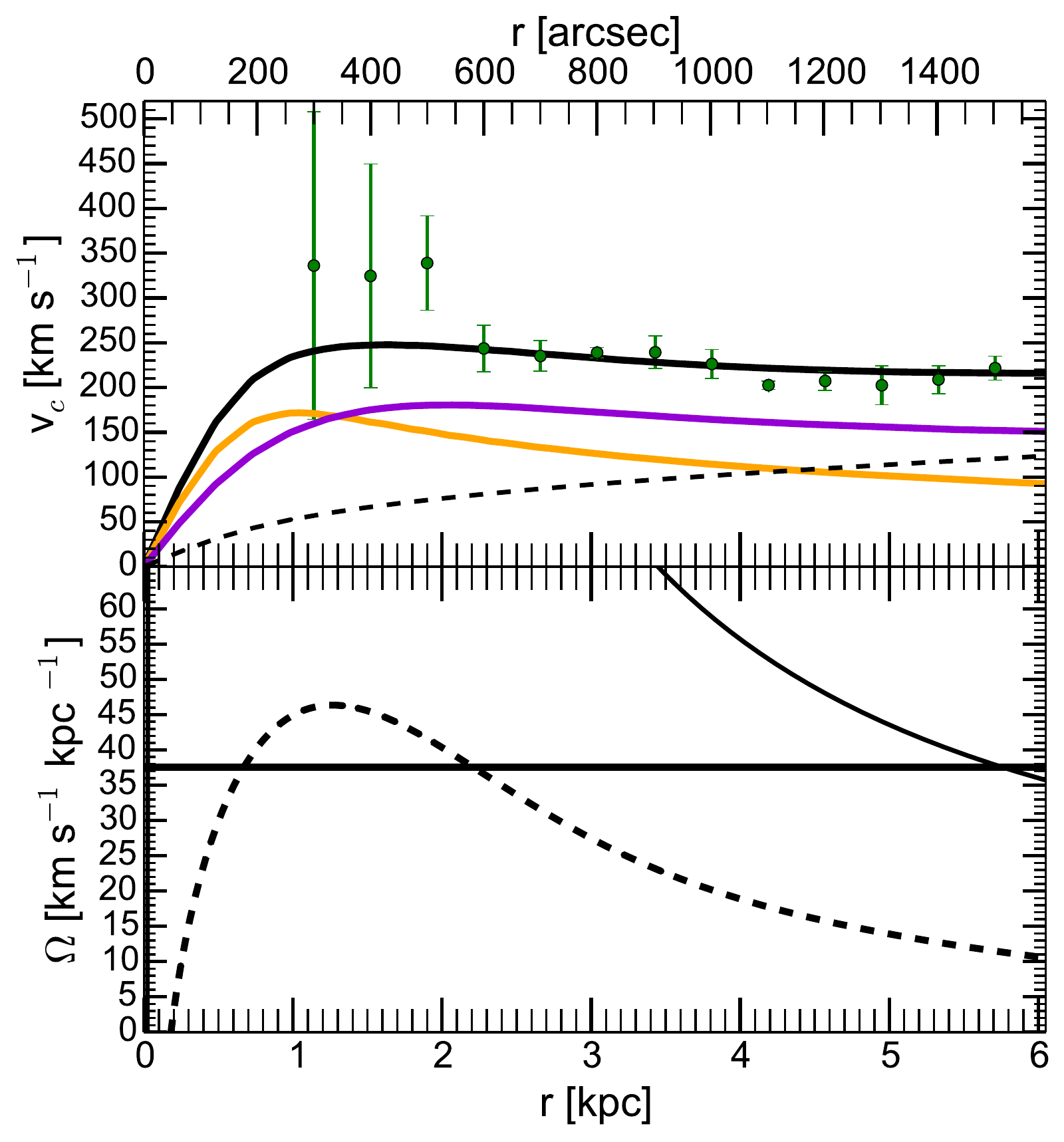}        
    \vspace{-0.4cm}
    \caption{
    \textit{Top panel}: Circular velocities of Model 1 (600\ut) and M31.
     ${\rm v}_{\rm c}$ curves of the different components of Model 1, the ICB (orange), the 
     disk + B/P bulge (purple), the dark matter (dashed curve) and the total 
    ${\rm v}_{\rm c}$ (solid black curve). 
    The rotation velocities estimated from HI observations (green dots) \citep{Chemin2009}. 
    \textit{Bottom panel}: The angular frequency profile ($\Omega$) of Model 1 at 600\ut (solid curve),
    and $\Omega_{\rm ILR}\e\Omega-\kappa/2$ (dashed curve).
    The pattern speed of the bar is $38\kms\kpc^{-1}$ (horizontal solid line), locating corotation 
    at $5.8\kpc$. The inner inner and the outer inner Lindblad resonances are located at $0.7\kpc$ and $2.2\kpc$ .
    }
    \label{fig:kinvcps}
  \end{center}
\end{figure}

In the top panel Fig.\ref{fig:kinvcps} we show the in plane-azimuthally averaged circular velocity 
(${\rm v}_{\rm c}$) curves of the different components of Model 1 and compare them with the rotation
curve estimated from HI observations \citep{Chemin2009}. 
The ${\rm v}_{\rm c}$ of the ICB component reaches a maximum circular velocity of 172\kms 
at \si1.0\kpc, dominating over the B/P bulge component 
within $R\si0.5\kpc$, and then drops rapidly. 
The B/P bulge component reaches a maximum of 180\kms at \si2.0\kpc, dominating over the ICB.
The total circular velocity shows a maximum of 248\kms at 1.6\kpc. 
Beyond 6\kpc the total ${\rm v}_{\rm c}$ of the model stays at around 210-220\kms,
slightly below the measured $v_{\rm c}\si240\kms$ \citep{Corbelli2010}. 
This could be remedied by adjusting the outer disk and halo mass 
distribution, but we do not attempt this in the present paper, focusing instead on the bulge.

In the bottom panel of Fig.\ref{fig:kinvcps} we show the angular frequency 
profile ($\Omega$) of Model 1. 
With the spatial and velocity scaling we calculate the pattern speed of Model 1, 
$\Omega_p\e0.29\uv\ud^{-1}$ in internal units, 
and $\Omega_p\e38\kms\kpc^{-1}$ in physical units.
The corotation radius, where $\Omega_p\e\Omega$, is located at $r_{\rm cor}\e5.8\kpc$.
We also calculate the inner inner and the outer inner Lindblad resonances $\Omega_{\rm ILR}\e\Omega-\kappa/2$ \citep{Lindblad1956}, 
obtaining $r_{\rm IILR}\e0.7\kpc$ and $r_{\rm OILR}\e2.2\kpc$. 
The outer Lindblad resonance $\Omega_{\rm OLR}\e\Omega+\kappa/2$  is located at $r_{\rm OLR}\e10.4\kpc$.
M31 shows a prominent ring-like structure at $\si10\kpc$ \citep{Habing1984, Haas1998, Gordon2006, Barmby2006} (10\kpc-ring), 
whose origin has been suggested to be due to: (1) a collision with another galaxy \citep{Block2006, Dierickx2014},
or (2) to a OLR resonance with the bar (AB06). 
If we assume that: i) the 10\kpc-ring is located near the OLR, and ii) that the circular velocity at OLR is $\si240\kms$ 
and roughly constant, we estimate a pattern speed of 
$\Omega_{\rm OLR}\e\Omega\,(1+1/\sqrt{2})\e240/10(1+1/\sqrt{2})\kms\kpc^{-1}\e\Omega_{\rm p}\e41.0\kms\kpc^{-1}$.
The fact that this is within 10 per cent of the value derived from the bulge structure 
suggests that the ring may indeed be related to the bar's OLR.

\subsubsection{Bulge mass profile}
\label{sec:res:bestmod:mass}
\begin{figure}
  \begin{center}
    \includegraphics[width=8.5cm]{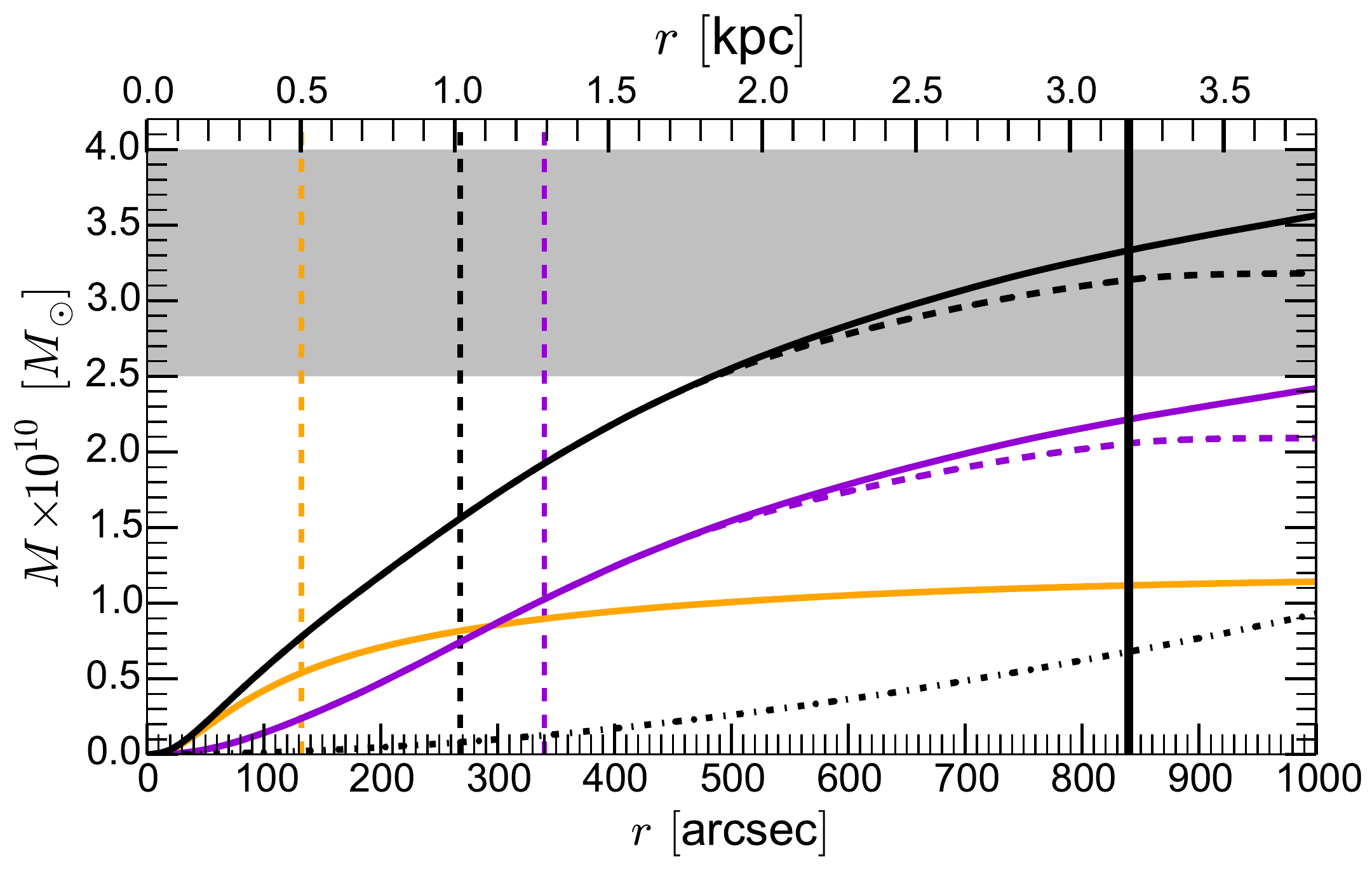}        
    \vspace{-0.8cm}
    \caption{
    Three-dimensional cumulative mass profiles $M\left(r\right)$ for Model 1 (600\ut) and its components: 
    ICB (orange curve), disk + B/P bulge (purple curve), disk + B/P bulge
    within a box-shaped volume (purple dashed curve), the combined bulges within spherical radius 
    (black solid curve), the combined bulges considering a box-shaped volume for the B/P bulge 
    (black dashed curve), and the dark matter halo (dash-dotted curve).
    The B/P bulge semi-major axis ends at 3.2\kpc (840\as) (solid vertical line).
    The shaded grey area marks the stellar mass for M31's bulge estimated by \citet{Kent1989} 
    (upper limit) and \citet{Widrow2005} (lower limit).
      The vertical dot-dash lines mark the deprojected half mass radius for each components.}
    \label{fig:mass}
  \end{center}
\end{figure}

In Fig.\ref{fig:mass} we show the three-dimensional cumulative radial mass profiles $M\left(r\right)$
within the bulge region. The spatial and velocity scaling gives a mass scaling of $1\um\e4.84\times10^{10}\sm$ for Model 1.
The initial mass distribution of the ICB is spherical and it changes in time only 
slightly, which makes its projected mass distribution almost independent of the spatial orientation.
In contrast to the ICB, it is not trivial to define the volume within which to measure the mass of a triaxial 
structure such as the B/P bulge. Therefore we make two estimations for the B/P bulge stellar mass, using  
two volumes. 
In the first estimation we consider all the mass within the B/P bulge 
spherical radius $r^{\rm B/P}\e3.2\kpc$ ($840\as$), obtaining a mass 
$M^{\rm B/P}\e2.2\times10^{10}\sm$. Within this radius the ICB component mass 
is $M^{\rm ICB}\e1.1\times10^{10}\sm$, giving a total stellar mass for the bulge of 
$M^{\rm Bulge}_{\rm Tot}\e3.3\times10^{10}\sm$.
If instead we consider the mass within a box\footnote{
The box major axis is $\Delta X\e6.4\kpc\,(1680\as)$, the minor axis $\Delta Y\e2.4\kpc\,(630\as)$ and vertical axis is 
$\Delta Z\e6.0\kpc\,(1580\as)$, which is perpendicular to the plane of the disk. 
$\Delta X$ major axis is defined based on the B/P bulge major axis 
$(2\times r^{\rm B/P})$. $\Delta Y$ is chosen to be the distance where the same isophote
that intersects the major axis $\Delta X$ in the face on view intersect the minor axis $\Delta Y$. 
$\Delta Z$ is chosen to be large enough to cover the whole boxy shape of B/P bulge.} 
we obtain $M^{\rm B/P}_{\rm Box}\e2.06\times10^{10}\sm$, \ie 6 per cent lower than in the spherical volume.
This difference is mostly due to the remaining material along the minor axis 
$\Delta Y$ between the disk and the B/P bulge. In both cases, the contribution to the 
total stellar mass of the bulge is $\si2/3$ for the B/P bulge and $\si1/3$ for the ICB.
The DMH mass within the B/P bulge radius is 
$M^{\rm DMH}(r^{\rm B/P})\e0.7\times10^{10}\sm$, which implies that within this 
radius the bulge components dominate the dynamics in this region, as is also shown by the circular 
velocity profiles in Fig.\ref{fig:kinvcps}.

The projected half mass radius of the ICB is $R_{\rm half}^{\rm ICB}\e0.4\kpc\,(100\as)$.
The deprojected half mass radius of the ICB and the B/P bulge ($M^{\rm B/P}$) are 
$r_{\rm half}^{\rm ICB}\e0.53\kpc\,(140\as)$ and $r_{\rm half}^{\rm B/P}\e1.3\kpc\,(340\as)$,
while the combined bulge half mass radius is $r_{\rm half}^{\rm Bulge}\e1.02\kpc\,(270\as)$.
We find that the ICB dominates within $r<265\as\!\approx\!1.0\kpc$ in the deprojected case, and 
$R\!<\!170\as\!\!\approx\!\!650\pc$ in the projected case. 
Beyond this transition region the mass of the B/P bulge dominates, reaching more than double 
the mass of the ICB component at the end of the B/P bulge.

The combined bulge stellar mass profile already reaches $M\left(r\right)\e2.5\times10^{10}\sm$ at $r\e470\as$, 
which is the mass estimated by \citet{Widrow2005} for M31's bulge, but it does not reach the mass 
estimate by \citet{Kent1989} of  $M\e4.0\times10^{10}\sm$. 
Using spectral energy distributions and rotation curves, \citet{Tamm2012} estimate even higher values for  
of M31 bulge mass (which include a stellar halo) ranging $(4.4-6.6)\times 10^{10}\sm$, while
\citet{Geehan2005} and \citet{Corbelli2010} estimate lower masses: $3.2\times 10^{10}\sm$ and $3.8\times 10^{10}\sm$,
respectively.
It is important to mention that most previous mass estimations for M31's bulge have assumed an
axisymmetric or oblate geometry for the bulge, where the mass of the disk strongly contributes within the bulge region.
In the models presented here all the stellar mass within the B/P bulge is considered as part of the bulge, 
its distribution is non-axisymmetric, and no separate massive disk component is present here 
(the B/P bulge is made from former disk material).

\subsection{The thin bar of M31}
\label{sec:res:spur}
Until now we have focused mostly on the bulge of M31, comparing 
it with our best model always at the 600\ut snapshot.
We now turn to the structure outside the B/P bulge, between $R\si500$ and 700\as in projection
(Fig.\ref{fig:SBmap:snapshot}, top panels), and try to determine the possible presence and 
properties of the thin bar, which in projection generates so-called \textit{spurs}. 
While our standard 600\ut snapshot is a good match to the main
properties of the B/P bulge as shown above, we describe in this
section an earlier snapshot at 500\ut which matches better
the isophotal properties of the thin bar in M31, because it has a thin bar 
shorter than the one at 600\ut.
Later we turn to the isophotal structures of M31 even further out, 
between 800\as and 1100\as, which we argue is material trailing the thin bar, that in 
projection reproduce the \textit{lobe-shaped} isophotes in M31 beyond its triaxial bulge.
As we mentioned in Section \ref{sec:meth:sim:bar}, we define the thin bar as the flat structure 
that is aligned with the B/P bulge, 
excluding additional transient material that could be attached to its ends, such as leading or trailing 
spiral arms.

\subsubsection{Length of the thin bar \& the spurs}
\label{sec:res:spur:thinbar}
\begin{figure*}
\centering
    \includegraphics[width=18cm]{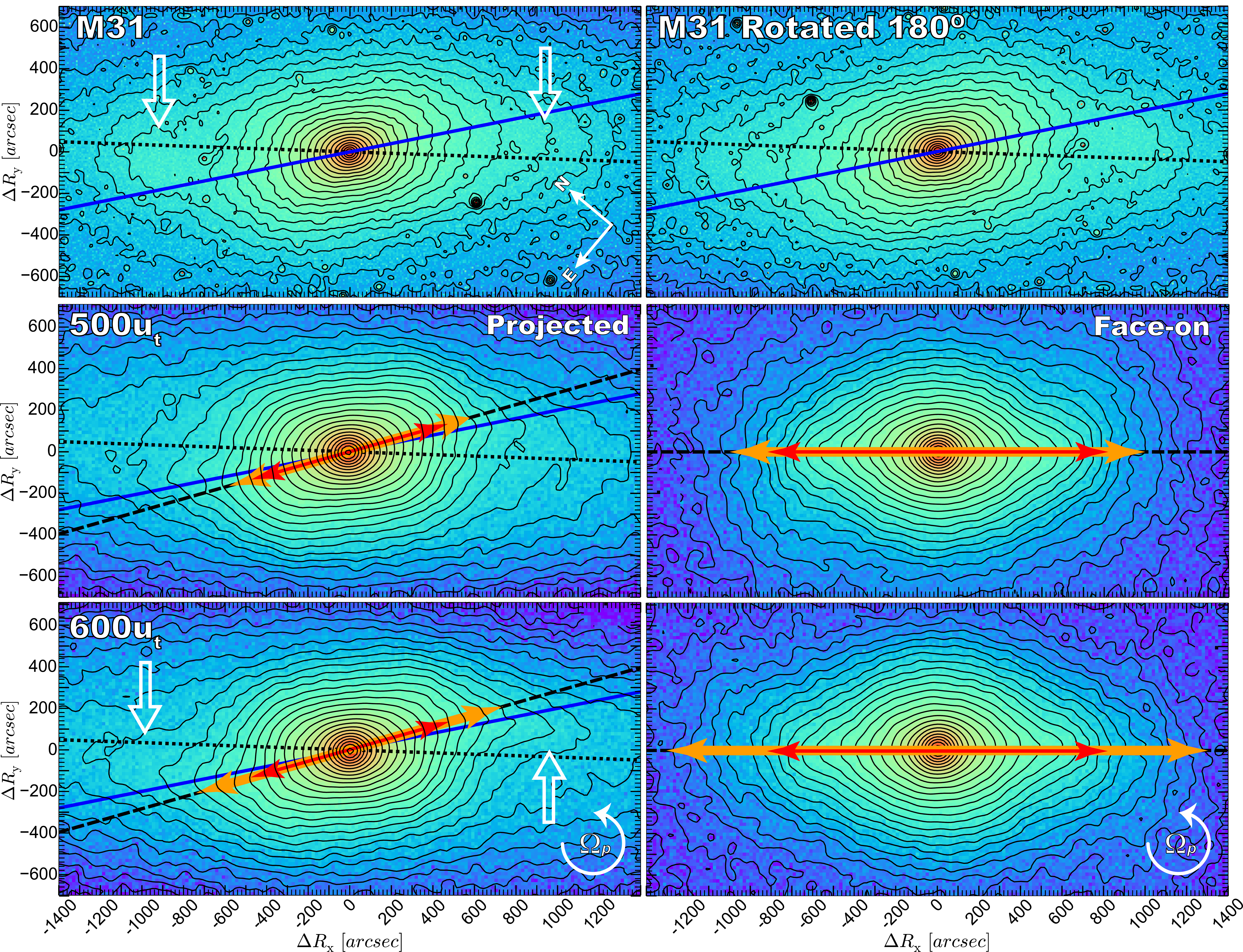}
     \vspace{-0.2cm}    
    \caption{
    SB map of M31 compared with two snapshots of Model 1 projected like M31 
    ($i\e77\degree$, $\theta_{\rm bar}\e54\degree\!\!.7$), and 
    face-on ($i\e0\degree$, $\theta_{\rm bar}\e0\degree$), showing the model's thin 
    bar extension (orange arrows) and B/P bulge extension (red arrows).
    \textit{Top panels}: 
    the left panel shows the M31 IRAC 3.6\mum image with the near side of the disk
    in the upper part of the image. We show ${\rm PA}_{\rm disk}\e38\degree$ (dotted line) and 
    ${\rm PA}^{\rm M31}_{\rm max}\e51\degree\!\!.3\pm1\degree\!\!.2$ (blue line).
    The position angle of the horizontal axis $\Delta R_{\rm x}$ is ${\rm PA}_{\rm h}\e40\degree$.
    The asymmetric lobe-shaped isophotes are at $\Delta R_{\rm x}\approx-1000\as$ 
    and at $900\as$ (white arrows). 
    The right panel shows the same image rotated 180\degree 
    to emphasize the asymmetry of the lobe-shaped structures. 
    \textit{Middle panels}:
    Model 1 at snapshot $500\ut$ in projection (left panel) and face-on (right panel) rotating 
    anticlockwise. The ${\rm PA}$ of the projected major axis of the bar is 
    ${\rm PA}_{\rm bar}\e55\degree\!\!.7$ (dashed line in left panels).
    The thin bar semi-major axis in the plane of the disk is $r_{\rm bar}^{\rm thin}\e4.0\kpc\,(1000\as)$
    (right panel). In 
    projection (left panel) the thin bar appears shorter, extending out only to
    $\Delta R_{\rm x}\e-580\as$ and 580\as (each end of the orange arrow).
    The B/P bulge semi-major axis is $r^{\rm B/P}\e3.2\kpc\,(840\as)$ (red arrow right panel) 
    and in projection extends only to $R^{\rm B/P}\e1.9\kpc\,(510\as)$ from the centre 
    (red arrow left panel). \textit{Bottom panels}: snapshot at $600\ut$ projected (left panel) and 
    face-on (right panel) rotating anticlockwise. The thin bar is longer than at $500\ut$, 
    with $r_{\rm bar}^{\rm thin}\e5.1\kpc\,(1300\as)$, extending in projection from 
    $\Delta R_{\rm x}\e-740\as$ to 740\as. The lobe-shaped isophotes are shown with white arrows. 
    Note that the face-on projections shown in these panels
    correspond to viewing the model from below relative to our line of sight to M31. Thus the rotation direction 
    indicated in the lower left panel corresponds to the sign of the velocities shown in Fig. \ref{fig:kinmap} 
    and to the projected sense of rotation shown in the lower left panel here (curved white arrows).}
    \label{fig:SBmap:snapshot}
\end{figure*}

In barred disk galaxies viewed at moderate inclination there are often 
isophotal elongations or so-called spurs extending outside their B/P bulges. 
These are also visible in N-body models of bars and are the natural projection 
of the thin bar outside the B/P bulge (ED13; see also AB06). 
Because they are thinner structures they lie closer to the projected 
angle of the major axis of the bar than the B/P bulge, which is vertically extended.

When the 600\ut snapshot of Model 1 is projected with $i\e77\degree$, $\theta_{\rm bar}\e54\degree\!\!.7$
and ${\rm PA}_{\rm disk}\e38\degree$ (note that the position angle of the horizontal axis $\Delta R_{\rm x}$ 
in Fig.\ref{fig:SBmap:snapshot} is ${\rm PA}_{\rm h}\e40\degree$), we also observe prominent spurs,
as shown in Fig.\ref{fig:SBmap:snapshot} (bottom panels). 
The thin bar semi-major axis in this snapshot is
$r_{\rm bar}^{\rm thin}\e5.1\kpc\,(1300\as)$ in the plane of the disk (bottom right panel).
We determine the value $r_{\rm bar}^{\rm thin}$ as the point where the ellipticity profile of the model viewed 
face on drops 15 per cent below its maximum value \citep{Martinez-Valpuesta2006}.
Approximating the thin bar as one dimensional locates in projection the ends of the thin bar semi-major axis at:
\begin{align}
\label{eq:bar}
&R_{\rm bar}^{\rm thin} \,\,= \pm r_{\rm bar}^{\rm thin}\,\left[\cos^2(\theta_{\rm bar})\!+\!\sin^2(\theta_{\rm bar})\cos^2(i)\right]^{1/2} \\
\label{eq:barxy}
&(\Delta R_{\rm x},\Delta R_{\rm y})= R_{\rm bar}^{\rm thin}\left(\cos({\rm PA}_{\rm bar} - {\rm PA}_{\rm h}),\sin({\rm PA}_{\rm bar} - {\rm PA}_{\rm h})\right)
\end{align}
where recalling Eq.\ref{eq:angle} and Eq. \ref{eq:PAbar} we have $\theta_{\rm proj}\e17\degree\!\!.7$ and 
the projected major axis of the thin bar at ${\rm PA}_{\rm bar}\e55\degree\!\!.7\pm2\degree\!\!.5$ 
(shown with orange arrows in Fig.\ref{fig:SBmap:snapshot}).
The extension of the thin bar semi-major axis is then $R_{\rm bar}^{\rm thin}\e\pm2.9\kpc\,(\pm770\as)$ 
roughly at $(\Delta R_{\rm x},\Delta R_{\rm y})\e(-740,-210)\as$ at the left side of the bulge, and at 
$(\Delta R_{\rm x},\Delta R_{\rm y})\e(740,210)\as$ at the right side of the bulge (bottom left panel).
These locations show the end of the thin bar in projection, and beyond this the spur shaped isophotes  
deviate from the projected bar axis, coming back to the disk major axis. 
The B/P bulge 3D semi-major axis is $r^{\rm B/P}\e3.2\kpc\,(840\as)$ and in projection extends only 
to $R^{\rm B/P}\e1.9\kpc\,(510\as)$ from the centre.

However, in M31 the isophotes in this region show no clear corresponding prominent 
features (Fig.\ref{fig:SBmap:snapshot} top panels). As explained in 
Section \ref{sec:res:sub:par:time} the snapshot at 600\ut was selected to 
fit the photometry in the bulge of M31. That it fails to match M31 outside 
this is not unexpected: bars are complex three dimensional structures and 
is not surprising that this diversity and complexity is not captured completely 
by a simulation.

To gain insight into which structures project to produce isophotes similar 
to M31 outside the B/P bulge, we use an earlier snapshot of the same model.
We do not claim that M31's thin bar necessarily went through a similar evolution 
between these two snapshots, but instead  we use the morphological 
structure and mass distribution of the thin bar at different snapshots as a tool 
to understand the present mass distribution and the morphology of the thin bar in M31.

In Fig.\ref{fig:SBmap:snapshot} (middle panels) is shown the earlier snapshot 
at 500\ut (3.8\Gyr), when the bar age is $t_{\rm bar}^{\rm age}\e2.32\Gyr$, 
and the B/P bulge age since the start of the buckling is $t_{\rm B/P}^{\rm age}\e1.5\Gyr$.
Using the same projection, the B/P bulge at this snapshot is similar to the B/P bulge at 
600\ut, but it manifests less prominent spurs, and is more similar to M31 in 
the spurs regions. This is the result of the thin bar being shorter, reaching only 
$r_{\rm bar}^{\rm thin}\e4.0\kpc\,(1000\as)$ from the centre (in the plane of the disk),
as shown in the face on view and therefore the total extension is $l_{\rm bar}^{\rm thin}\e8\kpc\, (2000\as)$. 
In projection the thin bar 
position angle is ${\rm PA}_{\rm bar}\e55\degree\!\!.7$ and it
extends to $R_{\rm bar}^{\rm thin}\e\pm 2.3\kpc\,(600\as)$.
This locates the ends of the projected thin bar at 
$(\Delta R_{\rm x},\Delta R_{\rm y})\e(-580,-160)\as$ and at 
$(580,160)\as$. Beyond these locations the spur shaped isophotes  
deviate from the projected bar axis similar
to the 600\ut snapshot, but closer to the centre.
A shorter thin bar has the consequence that the spurs 
are weaker and  resemble more closely the isophotes outside the M31 bulge at $\Delta R_{\rm x}\e-570\as$ 
and at 570\as, which we argue are the \textit{real} spurs generated by its thin bar.

Bars are classified as fast if they satisfy the criteria 
$\mathcal{R}\e r_{\rm cor}/r_{\rm bar}^{\rm thin}\!\leq\!1.4$ \citep{Debattista2000}. 
For Model 1 at 600\ut, the snapshot that best matches
the bulge of M31, the bar pattern speed is $\Omega_p\e38\kms\kpc^{-1}$, locating 
the corotation radius at $r_{\rm cor}\e5.8\kpc$, which combined with the thin bar semi-major 
axis length at 600\ut of $r_{\rm bar}^{\rm thin}\e5.1\kpc$ 
results in a ratio of $\mathcal{R}\e1.14$, \ie a fast bar.
In the 500\ut snapshot of Model 1, which best matches the spurs in M31, 
the bar pattern speed is slightly higher with $\Omega_p\e41\kms\kpc^{-1}$, 
placing the corotation radius at $r_{\rm cor}\e5.3\kpc$, which combined with 
the thin bar semi-major axis at 500\ut $r_{\rm bar}^{\rm thin}\e4.0\kpc$
results in a ratio of $\mathcal{R}\e1.32$ classifying this also as a fast bar.
While the 600\ut snapshot matches better the main properties of the B/P bulge, as shown in Table \ref{tab:param-time},
the 500\ut snapshot matches better the isophotal properties of the thin bar in M31.

\subsubsection{Material trailing the thin bar}
\label{sec:res:spur:lobes}
\begin{figure}
  \begin{center}
    \includegraphics[width=8.7cm]{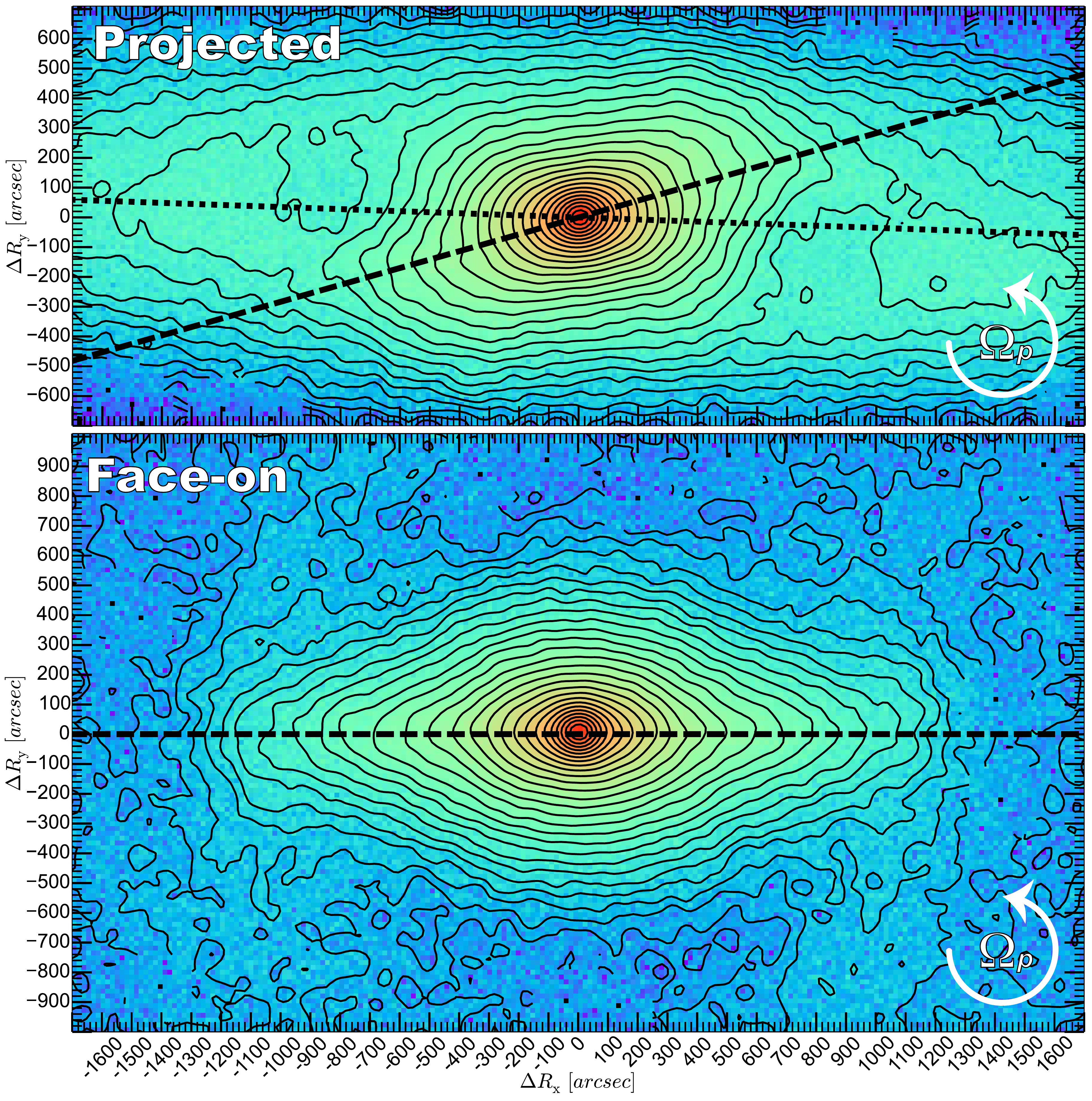} 
    \vspace{-0.7cm}    
    \caption{
    SB map of Model 4 at 800\ut rotating anticlockwise. 
    \textit{Top panel:} projection with $i\e77\degree$ and $\theta_{\rm bar}\e54\degree\!\!.7$,
     showing the disk major axis (dotted line) and the projected bar major axis (dashed line)
    with a position angle ${\rm PA}_{\rm bar}\e55\degree\!\!.7$. The near side of the disk 
    is in the upper part of the panel.
    \textit{Bottom panel:} face-on view ($i\e0\degree$, $\theta_{\rm bar}\e0\degree$), 
    with the bar major axis aligned with the horizontal axis (dashed line).}
    \label{fig:SBmap:mod4}
  \end{center}
\end{figure}

Now we focus on the structures of M31 located outside the B/P bulge and the thin 
bar regions. The top left panel in Fig.\ref{fig:SBmap:snapshot} shows 
that M31 has elongated isophotes with \textit{lobe-shaped} structures on both sides of 
the B/P bulge, which are asymmetric between each other. They are located at 
$\Delta R_{\rm x}\e-1000\as$ (left side of the bulge) and at $\Delta R_{\rm x}\e900\as$ 
(right side of the bulge), and they are very close to the disk major axis 
($\Delta R_{\rm y}\e\pm100\as$). To show more clearly the asymmetry we also include a rotated image of M31.

In comparison, Model 1 at 600\ut also presents such lobe-shaped structures
beyond the spurs region, as shown in Fig.\ref{fig:SBmap:snapshot} (bottom left panel)
at $(\Delta R_{\rm x},\Delta R_{\rm y})\e(-1100,-100)\as$ and at 
$(1200,100)\as$.
These features are generated by the material trailing the thin bar, as shown by 
the face on view (bottom right panel).
In the projected view (bottom left panel) we see that this material generates additional 
isophotal elongated structures beyond the spurs. They have the form of curved isophotes 
or \textit{lobes} at $\Delta R_{\rm x}\e-1000\as$ (left side of the bulge) and $1000\as$ 
(right side) that come back to the disk major axis, and are slightly asymmetric between 
the left and the right side. The snapshot at 500\ut also shows this features, but  
less prominently.

Such lobes are similar to the isophotes observed in M31 at 
$\Delta R_{\rm x}\e900\as$ (right side of the bulge) and even more similar to the 
side at $-1000\as$ (left side) as shown in Fig.\ref{fig:SBmap:snapshot} 
(top left panel). The notable asymmetry between the left side and the right side argues for 
the transient nature of these structures. 
Spiral arms like those seen near the M31 bar region \citep{Gordon2006, Barmby2006}, 
bar driven structures \citep{Martinez-Valpuesta2011}, or even the perturbation from a satellite 
passing near the centre \citep{Block2006, Dierickx2014}, could trigger transient structures that 
can change the shape of the isophotes around the thin bar.

AB06 argue that these elongated isophotes in M31 could be the projection of the thin bar, 
and find bar angles of $\theta_{\rm bar}\e20-30\degree$ by 
matching these structures with the projected thin bar of their N-body models. 
However, it would not then be possible to match the orientation of the B/P bulge 
without requiring a significant misalignment between the B/P bulge 
and the thin bar of more than 20\degree  in the plane of the disk. 
In our case we find the bar angle by matching the twist of the 
boxy bulge, and as a consequence the thin bar generates spurs further away from  
the disk major axis.

These \textit{lobe-shaped} structures are a common phenomenon in other galaxies, 
as shows by ED13 with some examples, such as NGC 4725 (their Fig.2), which present trailing spiral 
arms connected to its thin bar, generating also elongated isophotes that return to the disk 
major axis. In some galaxies the transient material may be observed in a leading configuration.
To demonstrate the effect of material trailing the bar more clearly we show Model 4 in Fig.\ref{fig:SBmap:mod4} with a
more unstable disk than the disk of Model 1. 
Model 4 was built with a constant $Q_{\rm T}\e1.0$ at the initial time and a slightly more massive
dark halo, obtaining a B/P bulge with a stronger peanut-shape. Fig.\ref{fig:SBmap:mod4} shows
the effect of transient spiral structures behind the bar on the projected 
isophotes at $\Delta R_{\rm x}\approx\pm1200\as$, connecting to the spurs of the thin bar at roughly 
$\Delta R_{\rm x}\approx\pm800\as$.
As our models are pure $N$-body systems, the disk is heated in time due to perturbations of the bar and/or the 
spiral activity, stabilising the disk and weakening in time the transient structures \citep{Sellwood2014a}. 
Only simulations of stellar disks with gas can remain locally unstable for longer times \citep{DOnghia2013}.

\section{Triaxial models for the bulge of M31 in the literature.}
\label{sec:bulgeM31}
Historically, the modelling of M31 bulge has been mostly made
with axisymmetric or oblate models \citep{Ruiz1976, Kent1989, Widrow2003, Widrow2005, Block2006, Hammer2010}. 
Only few models address the triaxiality of the M31 bulge,
either through an analytical approach or N-body simulations,
and they differ substantially in their properties.

Following the estimation of $\Delta {\rm PA}_{\rm max}\e10\degree$ by \citet{Lindblad1956}
between the bulge and the disk major axis, 
\citet{Stark1977} modelled M31 bulge three dimensional mass density as a triaxial 
ellipsoid with an apparent axial ratio of $q\e0.625$ and using $i\e77\degree$. He obtained  
a range of bar angles between $\theta_{\rm bar}\e25\degree\!\!.3$ and $86\degree\!\!.2$, with a
prolate solution at $38\degree$. 
\citet{Gerhard1986} found nearly-prolate solutions for a slightly larger projected axis ratio of $q\e0.85$,
obtaining a bar angle range between $\theta_{\rm bar}\e38\degree$ and $88\degree$, and
also finding a prolate solution at $38\degree$. If we use 
$\Delta {\rm PA}_{\rm max}\e\theta_{\rm proj}\e10\degree$ and Eq.\ref{eq:angle} for the 1-D bar approximation 
we obtain a similar result for the bar angle $\theta_{\rm bar}\e38\degree\!\!.1$.
To constrain the family of solutions, \citet{Stark1994} compared gas kinematic 
observations with pseudo-gas models using closed orbit analysis. The bulge was modelled as a 
spherical component plus a Ferrers bar with 30 per cent of the mass of the spherical component and 
$\Omega_{\rm p}\e57\kms\kpc ^{-1}$.  Comparing with the observed twist of the velocity 
iso-contours, they estimate bar angles between $\theta_{\rm bar}\e60\degree$ and $78\degree$.
\citet{Berman2001} and \citet{Berman2002} explored full 2D hydrodynamical simulations
using a static potential for M31's stellar and dark matter component, with a triaxial
component for the bulge, and compared them with CO observations, estimating a bar 
angle of $\theta_{\rm bar}\e15\degree$ and a pattern speed of 
$\Omega_{\rm p}\e51$ -- $55\kms\kpc ^{-1}$.

Besides the present paper, only AB06 have compared full N-body simulations of 
B/P bulges combined with classical bulges with the isophotes of M31's bulge. 
They used two pure B/P bulge models, where one was a weak boxy bulge, and the other 
a strong X-shaped bulge. They also used two models that combined classical bulges with
B/P bulges. They excluded the extreme X-shaped pure B/P bulge model, 
because it resulted in isophotes that are too pinched in the bulge region. 
They also excluded the weak pure B/P bulge due to its weak spurs.
This left the models with composite bulges as their best candidates, and they concluded 
that M31 is likely to have a classical bulge component as well as a B/P bulge.

Here we also find a solution with a composite bulge, but with three major 
differences. The first is our estimated bar angle. While AB06 find 
angles between $\theta_{\rm bar}\e20\degree$ -- $30\degree$ trying to match the lobe-shaped
structures in M31 with the projection of the thin bar of their N-body models, we estimate an angle 
of $\theta_{\rm bar}\e54\degree\!\!.7$ that reproduces the isophotal twist 
of M31's bulge. This would generate a misalignment of more than 20\degree between the thin 
bar major axis and the B/P bulge major axis. Instead, here we find a solution where no misalignment 
is necessary.

The second difference are the thin bar properties. With the 
estimated bar angle $\theta_{\rm bar}$ the position angle of projection of the thin bar 
major axis is ${\rm PA}_{\rm bar}\e55\degree\!\!.7$. Also, 
according to our comparison with the 500\ut snapshot of Model 1,
the estimated length for the thin bar 
semi-major axis that matches better the weak spurs observed in M31 is 
 \si4\kpc (1000\as) in the plane of the disk, instead of the estimated 1320\as of AB06.
If the thin bar were much longer than 1000\as, it would generate spurs much 
more prominent than observed, as we showed with the thin 
bar at 600\ut which has a semi-major axis of 5.1\kpc (1300\as) in the plane of the disk.
 
Finally, the properties of the classical bulge are different. We find a 
massive, but concentrated ICB component, that combined with the B/P bulge reproduces the
surface-brightness profile and the morphology of M31's bulge, and exclude less concentrated solutions for the ICB. 
The Hernquist ICB mass ranges explored by AB06 are similar to our massive models in the ICB 
parameter space exploration, but they considered larger scale lengths, with 
$r_{\rm b}\e0.4\ud$ and $0.6\ud$. As shown in Fig. \ref{fig:MHRH}, this range 
of scale length results in S\'ersic indices lower than is required to match M31. 
Therefore the surface brightness profile requires a concentrated classical 
bulge.

\section{Conclusions}
\label{sec:conc}
We have presented here a dynamical model that reproduces the main photometric 
observables of the bulge of the Andromeda galaxy.
We explored a large set of N-body models, combining B/P bulges with classical bulges. The 
B/P bulges are generated in the simulations from the initial disk that naturally forms a bar and buckles generating the 
boxy structure which evolves together with the ICB, resulting at the end in a system in dynamical equilibrium.
We specially focus on exploring the size and mass of the classical bulge component.

By a quantitative comparison of morphological and kinematic properties of M31 with the models, 
we are able to find a best model for the bulge of M31. This model requires a classical bulge and a B/P bulge 
with masses of 1/3 and 2/3 of the total stellar mass of the bulge to match the observations.
The classical bulge contributes mainly in the centre of the bulge, within $\si530\pc\,(140\as)$, 
increasing the mass concentration and therefore the S\'ersic index. The cuspy density profile is also reflected in 
the kinematics, generating a dispersion drop in the centre. On the contrary, the mass contribution 
of the B/P bulge in the centre is shallow which lowers the S\'ersic index of the combined SB profile. 
Beyond $\si530\pc\,(140\as)$ the B/P bulge dominates, explains the observed rotation, and the 
boxy shape of the isophotes. We excluded pure B/P bulge models, because they show a S\'ersic 
index too low to reproduce the value observed in M31, and because their central velocity dispersion 
lacks the drop generated by the ICB component.

While our standard model is a good match to the main properties of the B/P bulge, we
find an earlier snapshot of the same model better matches
the isophotal properties of the thin bar in M31. 
Comparing this snapshot to the weak spurs observed in M31 at the end of its boxy bulge
suggests that the thin bar is short, with a semi-major axis of 
$r_{\rm bar}^{\rm thin}\e4.0\kpc\,(1000\as)$ in the plane of the disk, and 
in projection extends to $R_{\rm bar}^{\rm thin}\e2.3\kpc\, (600\as)$ located at 
a position angle of ${\rm PA}_{\rm bar}\e55\degree\!\!.7$. 
M31 also shows lobe-shaped isophotes further away at $R\si3.4\kpc\, (950\as)$ located 
near the disk major axis. The proximity of these lobe-shaped isophotes to the disk major axis
and the asymmetry between the structures at both sides of the bulge suggests the presence 
of transient structures trailing the thin bar.

The presence of a massive B/P bulge component intertwined with a classical bulge has strong 
implications for the formation and the secular evolution history of M31. To better understand and quantify the 
impact of the B/P bulge on the dynamics of the galaxy, we are developing Made-to-Measure 
models that reproduce simultaneously the thin bar and B/P bulge structures, and also 
M31's disk mass distribution (Bla\~na et al. in prep.) using detailed IFU kinematics 
(Opitsch et al. in preparation).

\section*{Acknowledgements}
\label{sec:acknow}
Matias Bla\~{n}a (MB) would like to thank Manuel Behrendt, Alessandro Ballone and 
Fabrizio Finozzi for several insightful scientific discussions,
and thank also Achim Bohnet for the technical support.
We are greatful to Pauline Barmby for providing the Spitzer 3.6\mum IRAC1 data.
We also thank the anonymous referee for constructive comments that improved the manuscript.
MB would also like to thank the Deutscher Akademischer Austauschdienst (DAAD) 
for the support of this project with its Research Grant
for Doctoral Candidates and Young Academics and Scientists (57076385). 

\bibliographystyle{mnras}

\appendix
\section{Initial Conditions and Bar Formation}
\label{sec:appA}
We modified {\sc MaGalie} following the procedure of \citet{Athanassoula2002} to generate galaxies 
with a constant Toomre parameter $Q_{\rm T}$ \citep{Toomre1964}. 
This allows setting disks that are dynamically colder at all radii, 
than an exponential radial velocity dispersion profile. This is in principle better 
in the inner regions of the disk, because it respects the epicyclic approximation 
used by {\sc MaGalie} at that radii. A colder disk also is more bar-unstable.
As an example, the solutions of the solid body rotator disk \citep{ Kalnajs1965}
show that these systems are unstable to bar modes when the ratio ($q_{\rm rot}$) of 
rotational kinetic energy to potential energy is $q_{\rm rot}\!>\!0.1286$ \citep{Binney2008}. 
Making a disk colder favours bar formation (or in general non-axisymmetric instabilities), 
because this distributes more kinetic energy in the rotational component of the disk particles
than in the random motion component, increasing $q_{\rm rot}$. We achieve this by choosing a radial velocity 
dispersion of the form $\sigma^{Q_{\rm T}}_r\e3.36\,Q_{\rm T}\,\Sigma\, \kappa ^{-1}$,
where $\kappa$ is the epicyclic frequency, $\Sigma$ the surface mass density and $Q_{\rm T}$ 
is the initial value of the Toomre parameter. We choose an initial value of $Q_{\rm T}\e1.0$ to 
avoid axisymmetric instabilities, although depending on the disk thickness used, this limit can be 
as low as $Q_{\rm T}\e0.696$, as shown by \citet{Behrendt2015}. 

In some extreme cases the circular velocity can be very low if the 
DMH has a low concentration or there is no initial bulge component. This forces the Jeans equations 
to assign streaming or azimuthal velocities to the particles that would be higher than the 
circular velocity within a small radius  $r_{\rm C}$. We prevent this by changing the radial velocity 
dispersion within $r_{\rm C}$ by a profile that also respects the epicyclic approximation and has the 
form $\sigma^{\Sigma}_r\e C\,\Sigma^{-1/2}$, where $C$ is a constant determined at $r_{\rm C}$ 
to make a continuous dispersion profile, i.e. $\sigma^{Q_{\rm T}}_r(r_{\rm C})\e\sigma^{\Sigma}_r(r_{\rm C})$. 
\label{lastpage}
\end{document}